\documentclass[10pt,journal]{IEEEtran}
\newtheorem{theorem}{Theorem}

\newtheorem{lemma}{Lemma}
\newtheorem{definition}{Definition}
\newtheorem{remark}{Remark}

\newtheorem{assumption}{Assumption}
\newtheorem{corollary}{Corollary}

\usepackage{graphicx}
\usepackage{subfigure}
\usepackage{array} 
\usepackage{amsmath}
\usepackage{amssymb}
\usepackage{mathtools}        
\usepackage{latexsym}
\usepackage{mathrsfs}
\usepackage{cite}
\usepackage{enumerate} 
\usepackage{float}
\usepackage{amsfonts} 
\usepackage{epsfig} 
\usepackage{epstopdf}
\usepackage[dvipsnames]{xcolor}
\usepackage{tikz}\newcommand*{\circled}[1]{\lower.7ex\hbox{\tikz\draw (0pt, 0pt)%
		circle (.4em) node {\makebox[1em][c]{\small #1}};}}
\usepackage{color}
\usepackage[scaled]{helvet}
\usepackage{tikz,xcolor,hyperref}

\begin{document}

\title{Robust Adaptive Prescribed-Time Control for Parameter-Varying Nonlinear Systems}
\author{Hefu Ye,  and Yongduan Song$^*$, \IEEEmembership{Fellow, IEEE}
	\thanks{This work was supported by the National Natural Science Foundation of China under grant (No.61991400, No.61991403, No.61860206008, and No.61933012). (Corresponding Author: Yongduan Song.)}
	\thanks{H.  Ye  and Y. Song are   with  the School of Automation, Chongqing University,  China.    (e-mail:    yehefu@cqu.edu.cn; ydsong@cqu.edu.cn).} 
}
\maketitle

\begin{abstract}
It is an interesting open problem to achieve adaptive prescribed-time control for strict-feedback systems with unknown and fast or even abrupt time-varying parameters. In this paper we present a solution with the aid of several design and analysis innovations. First, by using a spatiotemporal transformation, we convert the original system operational over finite time interval into one operational over infinite time interval, allowing for Lyapunov asymptotic design and recasting prescribed-time stabilization on finite time domain into asymptotic stabilization on infinite time domain. Second, to deal with time-varying parameters with unknown variation boundaries, we use \textit{congelation of variables} method and establish  three separate adaptive laws for parameter estimation (two for the unknown parameters in the feedback path and one for the unknown parameter in the input path), in doing so we utilize two tuning functions to eliminate over-parametrization. Third, to achieve asymptotic convergence for the transformed system, we make use of nonlinear damping design and non-regressor-based design  to cope with time-varying perturbations, and finally, we derive the prescribed-time control scheme from the asymptotic controller via inverse temporal-scale transformation. The boundedness of all closed-loop signals and control input is proved rigorously through Lyapunov analysis, squeeze theorem, and two novel lemmas built upon the method of \textit{variation of constants}. Numerical simulation verifies the effectiveness of the proposed method.   
\end{abstract}


\section{Introduction} 
\IEEEPARstart{P}rescribed-time control originated and motivated from the field of missile guidance has gained increasing attention since the seminal work\cite{2017-song-prescribed-time}.   In earlier studies of missile guidance and control, the well-known proportional navigation feedback method was used to regulate the system output to the target point in a prescribed time, regardless of the initial condition and any other design parameter\cite{1973-later-optimal}--\hspace{-0.08pt}\cite{Zarchan07}. However, this method was only applied to simple models such as double integrators. It is indeed nontrivial to achieve prescribed-time stabilization for high-order nonlinear systems, which is particularly true in the context of the system dynamics involving unknown time-varying uncertainties.

The first systematic approach addressing prescribed-time control of high-order nonlinear systems in normal form is proposed in \cite{2017-song-prescribed-time} based on  a novel time-varying state-scale transformation. Recently, the work \cite{2020-chitour-stabilization} revisits the problem of prescribed-time stabilization for such system using proportional navigation feedback and constructs a new prescribed-time controller in a more straightforward way, while giving the appropriate control gain by solving the linear matrix  inequality, thus eliminating the need for the small-gain theorem as in \cite{2017-song-prescribed-time}.  Thereafter, for such systems, a prescribed-time stabilizer with decreasing linear function constraints is investigated in\cite{2021-shakouri-prescribed-linear-decay}, where Stirling numbers and matrices are applied for the first time to prescribed-time control design and stability analysis. Results for prescribed-time stabilization for other types of systems are successfully established in subsequent studies, including time-delay systems\cite{2021-Espitia-TAC-PTC}--\hspace{-0.08pt}\cite{2022-espitia-sensor}, stochastic nonlinear systems\cite{2022-Li-prescribed-SSC}--\hspace{-0.08pt}\cite{2021-liwuquan-stochastic-prescribed}, Euler-Lagrange systems\cite{2021-yilang-EL-1,2022-ye-tac}, $p$-normal nonlinear systems\cite{2022-p-normal}, multi-agent systems\cite{2018-yucelen-finite,2020-yucelen-time-scaling},  strict-feedback-like systems\cite{2020-NYU-dynamic-high-gain}, and standard strict-feedback systems without/with unknown control gains\cite{2021-TAC-huachanghun-prescribed-adaptive,2022-hua-Cyber}, etc. It is worth emphasizing again that the settling time of the controlled system in the aforementioned works can be pre-set by users freely, irrespective of initial condition and any other design parameter. Therefore, prescribed-time control has its unique advantages over finite-time control\cite{2005-Orlov} and fixed-time control\cite{2020Polyakov}, since the settling time obtained by the later methods always depends on the initial condition and/or design parameters.
In \cite{2018-Sanchez-predefined-time} and \cite{2020-San-TAC}, the predefined-time control design and the Lyapunov-like conditions of predefined-time stability are investigated, respectively. Compared to prescribed-time control, such methods are effective in mitigating the effects of measurement noises. However, prescribed-time control is preferable to predefined-time control in some respects. For example, the former has smooth rather than discontinuous control inputs; the former can arbitrarily set an exact convergence time rather than its upper bound;  and the control effort of the later increases exponentially with the feedback signal whereas the former does not. Moreover, it is not clear how to extend the method of predefined-time control to systems with unknown high-frequency input gains.


On the other hand, modeling uncertainties arisen from unknown or even fast time-varying parameters are inevitable in practice. For example, in flight vehicle with propulsion actuation, the total mass of the vehicle decreases during the system operation. Some efforts have been made in adaptive control design for systems with unknown and time-varying parameters \cite{chenkaiwen,chenkaiwenIFAC,yehefu}. However, there is still no result for adaptive prescribed-time control for such systems.

Motivated by the above analysis, here in this work we propose adaptive prescribed-time control schemes for strict-feedback nonlinear systems with fast time-varying parameters in both the feedback and input paths. In our development, we make use of nonlinear damping design\cite{khalil}, adaptive backstepping design\cite{krstic}, and temporal-scale transformation based design\cite{2020-yucelen-time-scaling} as well as the \textit{congelation of variables} method\cite{chenkaiwen}. Several design and analysis innovations are needed in utilizing those methods for adaptive prescribed-time control. First of all,  the model considered in this work is more general and more challenging than that in \cite{khalil,krstic,2020-yucelen-time-scaling,2021-TAC-huachanghun-prescribed-adaptive} since the unknown parameters are allowed to be fast or even abrupt time-varying and the resultant uncertainties are essentially mismatched. 
Additionally, to completely remove the restrictive condition that requires the bounds on the radius of $\Theta_0$ (where $\Theta_0$ represents some compact sets related to $\theta(t)$) in \cite{chenkaiwen} be known a priori, we propose to use  a two-level estimation for time-varying parameters $\theta(t)$. More importantly, to cope with the uncertainties caused by time-varying perturbations and unknown control coefficients to achieve zero-error convergence, instead of using complex matrices known as regressors in control design and parameter estimator design, we resort to non-regressor-based approach for designing negative feedback to make the transformed system stable. The immediate benefit gained from such treatment is that, for high-order nonlinear systems in normal-form, a filter variable can be utilized to alleviate the computational burden in backstepping design, resulting in simplified control algorithms and facilitating the real time  implementation. The contribution of this paper is threefold:
\begin{itemize}	 
	\item For nonlinear systems with time-varying parameters in the feedback path and the input path, we propose a unified control framework that  achieves  asymptotic, exponential, super-exponential and prescribed-time stability, and these results can be established by selecting different design parameters without the need for alternating the entire controller structure;  
	\item Unlike those time-scale transformation-based methods that requires \textit{a priori} knowledge of the control coefficients (see, for instance, \cite{2018-yucelen-finite,2020-NYU-dynamic-high-gain,2020-yucelen-time-scaling}), the proposed method for the first time solves the prescribed-time stabilization for parameter-varying systems with both mismatched uncertainties and unknown control coefficients, providing a feasible control solution for a broader class of systems; 
	\item  The proposed strategy does not involve complex computation for regressor matrix and is based upon a relaxed condition on the unknown time-varying parameters of the feedback path, which is in contrast to the existing works\cite{chenkaiwen,chenkaiwenIFAC,yehefu}. For systems with time-varying parameters in normal form, the proposed control solution becomes simple in structure and  inexpensive in computation. 	
\end{itemize}

The remainder of this article is organized as follows. We begin our problem statement in   Section  \ref{problem-statement} with SISO parametric strict-feedback systems involved time-varying parameters both in the feedback path and the input path, followed by three useful lemmas  in Section \ref{Useful Lemmas}. The control design that incorporates four different steps is presented in Section \ref{Control Design}, which are, system reparameterization, Lyapunov design by means of congealed variables, Lyapunov redesign with tuning functions, and negative feedback control gain design. Section \ref{stability-analysis} addresses the stability analysis by considering the convergence of system states and  the boundedness of control input and update laws. To verify the effectiveness and benefits of the control algorithms, simulations on a benchmark example and  on the model of the ``wing-rock" unstable motion in high-performance aircraft at high angle of attack are performed and the results are given in Section \ref{Simulation}. The article is closed in Section \ref{CONCLUSIONS}. 

\textit{Notations:}   $\mathbb{R}$ is the field of reals, $\mathbb{R}_+=\{a\in\mathbb{R}: a >0\}$ and $\mathbb{R}_{\geq0}=\{a\in\mathbb{R}: a\geq0\}$.  $\mathbb{R}^n$ denotes the $n$-dimensional Euclidean space, 	$\mathbb{R}^{n\times m}$ is the set of $ n \times m$ real matrices.  $\Gamma \succ 0$  means that the symmetric matrix $\Gamma$ with suitable dimensions   is positive definite.   $f^n$ denotes the $n$th power of $f$ and  $f^{(n)}  =\frac{d  ^n  }{dt^n}f $ denotes the $n$th  derivative \textit{w.r.t.} $t$  of $f $.	 $f^{-1}(\cdot)$ is the inverse of a function $f(\cdot)$.  $f\in\mathcal{C}^{\infty}$ denotes a function  $f$ has  continuous derivatives of order $\infty$.   $w^{\top}$ and $\|w\|$ denote  the transpose  and the Euclidean norm of the vector $w$, respectively. A function $\xi:[0,\infty)\rightarrow [0,\infty)$ belong to the class  $\mathcal{K}$ if $\xi(0)=0$ and $\xi$ is increasing, i.e., $t_1<t_2\Rightarrow\xi(t_1)\leq\xi(t_2)$.  A continuous function $\varrho(t_1,t_2):\mathbb{R}_{\geq0}\times\mathbb{R}_{\geq0}\rightarrow\mathbb{R}_{\geq0}$ belongs to the class $\mathcal{KL}$ if $\varrho(\cdot, t_2)\in\mathcal{K}$ for any fixed $t_2$ and $\varrho(t_1,\cdot)$ is decreasing to zero for any fixed $t_1$.  $\lim_{t \rightarrow\infty} f(\cdot)$ denotes the limit of $ 	f(\cdot)$ as $ t \rightarrow \infty.$  $\breve{x}(\tau)$ (or $\breve{x}$) and $x(t)$ (or $x$) refer to  the value of the same signal on different time axes.   Both $e^{t}$ and $\exp(t)$ denote the exponential function.


\section{Problem Formulation and Some Preliminaries}\label{preliminaries}
\subsection{Problem Formulation}\label{problem-statement}
Consider the following strict-feedback systems with unknown time-varying parameters:
\begin{equation}\label{system}
\left\{\begin{array}{ll}
\dot{x}_i=\phi_i^{\top}(\underline {x}_i)\theta(t)+x_{i+1},~i=1,\cdots,n-1\\  
\dot{x}_n=\phi_n^{\top}(\underline{x}_n)\theta(t)+b(t)u,
\end{array}\right.
\end{equation}
where $\underline{x}_i=[x_1,\cdots,x_i]^{\top}\in \mathbb{R}^i$ is the state vector, $u\in\mathbb{R}$ is the input. The functions $\phi_i:\mathbb{R}^i\rightarrow\mathbb{R}^q,~i=1,\cdots,n,$ are smooth   and satisfy $\phi_i(0)=0.$  
The system parameters  $\theta(t)\in\mathbb{R}^q$ and $b(t)\in\mathbb{R}$ are unknown and time-varying and satisfy the following assumptions: 

\begin{assumption} \label{assumption1}
	The parameter $\theta(t)$ is piecewise continuous and $\theta(t)\in \Theta_0,$ for all $t\geq 0$, where $\Theta_0$ is a completely unknown compact set.   The ``radius" of $\Theta_0$, denoted by $\delta_{{\theta}}$, is  also unknown.
\end{assumption}
\begin{remark}
	The condition imposed on the parameters on the feedback path as in Assumption \ref{assumption1} makes the model more general than the one considered in \cite{2021-TAC-huachanghun-prescribed-adaptive}, since the latter requires that $\theta(t)$ be time-invariant. It is also makes the model more general than the one considered in \cite{chenkaiwen} because the latter requires that $\delta_{{\theta}}$ be known. 
\end{remark}

\begin{assumption} \label{assumption2}
	The control direction is known and does not change. We assume that  $b(t)$  is unknown but bounded away from zero in the sense that there exists an unknown constant $\ell_b$, such that   $0< |\ell_b| \leq  |b(t)| $, for all $t\geq 0$. In addition, there exists a known constant $\underline{b}$ such that $\underline{b}\leq |b(t)|.$
\end{assumption}

\begin{remark}	
	Assumption  \ref{assumption2} is a mild variation of the one imposed in \cite{chenkaiwen}. This assumption allows the systems considered in \cite{2020-NYU-dynamic-high-gain
		,2021-zhoubin-tac-prescribed,2021-TAC-huachanghun-prescribed-adaptive} to be treated as special cases wherein a priori precise knowledge on the control coefficients is needed, while here only crude information on the gain $b(t)$ is required. 
\end{remark}


In this study, we assume that all the system states are available for control design (for the case that only partial states are available, observer is needed, which, however, is beyond the scope of this work).

\begin{definition}[{[\citen{2021-Espitia-TAC-PTC}]}]\label{definition1} 
	The origin of the system $\dot{x}=f(x,t)$ is said to be  prescribed-time globally stable  in time $T$ if there exist a class $\mathcal{K} \mathcal{L}$ function $\varrho$ and a function $\mu:\left[0, T\right) \rightarrow \mathbb{R}_{\geq 0}$ such that $\mu$ tends to $\infty$ as $t$ goes to $T$ and, $\forall t \in\left[0,  T\right)$
	$$
	\|x(t)\| \leq \varrho\left(\left\|x\left(0\right)\right\|, \mu\left(t \right)  \right),
	$$
	where $x\in\mathbb{R}^n$ and $T$ is the settling time that can be prescribed in the design. 
\end{definition}
\subsubsection*{\textbf{Control Objective}} The control objective in this paper is to design  state-feedback adaptive  control schemes with bounded control inputs and bounded parameter estimations for system (\ref{system}) to achieve prescribed-time stabilization in the sense of Definition 1.



\subsection{Useful Lemmas}\label{Useful Lemmas}
A key ingredient in the stability analysis for prescribed-time stabilization is to ensure the boundedness of control input, which usually relies on proper control structure and suitable control gains. In particular, \cite{2021-TAC-huachanghun-prescribed-adaptive} requires that a number of gains must be selected to be larger than some constants associated with the system dimension. In \cite{2017-song-prescribed-time}, the small-gain and the ISS argument are adopted to select suitable control gains. Furthermore, in  \cite{2020-chitour-stabilization}, \cite{2021-shakouri-prescribed-linear-decay}, and \cite{2021-zhoubin-tac-prescribed}, the control gain is associated with the solution of a linear matrix inequality, a Lyapunov equation, and a parametric Lyapunov equation, respectively. Different from the aforementioned works, we here introduce a novel lemma to provide guidance for choosing  control gains properly, which is critical in the stability analysis in Section \ref{stability-analysis}.
\begin{lemma}\label{lemma1}
	Suppose there exists  a $\mathcal{C}^{\infty}$  positive function $\gamma$ satisfying $\lim_{\tau\rightarrow\infty}\gamma(\tau)=\infty$ and let $k_i$ be a positive constant defined by  $k_i=\inf\{\mathsf{K}_i(\tau)\}$.   For $\forall \tau\in[0,\infty)$ and $\forall \breve z_i \in \mathcal{L}_\infty$, if the following conditions hold:
	\begin{subequations} 
		\begin{align} 
		&(i)~~~~~~ \frac{d \breve z_i }{d\tau}=-\mathsf{K}_i(\tau) \breve z_i+\frac{1}{\gamma^\sigma}\mathcal{Y}_i(\breve z_i,\tau),~~\sigma=1,2,\cdots,n  \label{condition1}\\
		&(ii)~~~~  	\lim_{\tau\rightarrow\infty} {e^{-k_i\tau}} {\gamma^\sigma}=0,    \label{condition3}\\
		&(iii)~~~ 	\lim_{\tau\rightarrow\infty}\breve z_i=0,   \label{condition2}\\
		&(iv)~~~ \lim_{\tau\rightarrow\infty}\left(k_i-\frac{\sigma}{\gamma}\frac{d\gamma}{d\tau }\right)>0,\label{condition4} 
		\end{align} 
	\end{subequations}
	where $\mathcal{Y}_i(\breve{z}_i,\tau)\in\mathcal{C}^{\infty}$  satisfies $\mathcal{Y}(0,\tau)=0,$ then, 
	\begin{equation}
	\breve z_i\gamma^\sigma(\tau) \rightarrow 0~~\text{as}~~\tau\rightarrow \infty.
	\end{equation}
\end{lemma}
\textit{Proof:} see Appendix I.  $\hfill\blacksquare $

\begin{corollary}\label{corollary}
	Under the conditions of Lemma \ref{lemma1}, if $\mathsf{K}_i(\tau)\equiv k_{\min}$ is a positive constant, then the conclusion in Lemma \ref{lemma1} also holds.   
\end{corollary}
\textit{Proof:}  The proof is omitted as it is subsumed in the proof 
of Lemma \ref{lemma1}. $\hfill\blacksquare$

\begin{remark}\label{gamma(t)}
	By choosing $k_{i}>1$ and a suitable $\gamma(\tau)$, conditions (\ref{condition3}) and (\ref{condition4}) are easily satisfied. In addition, two $\gamma(\tau)$-type functions can be  directly derived from  [\citen{2020-NYU-dynamic-high-gain}, Remark 3],   and [\citen{2020-yucelen-time-scaling}, Remark 1], as follows:
	\begin{equation}
	\gamma(\tau)={a_0}\left(\frac{\tau}{a_0T}+1\right)^{2},~~\gamma(\tau)=\frac{1}{T} e^{\tau}.
	\end{equation}
	where $a_0$ and $T$ are positive constants.
\end{remark}

\begin{lemma}\label{lemma-s}
	Consider a special case of system (\ref{system}) in the absence of mismatched uncertainties, namely $\{\phi_i(\underline{x}_i)\}_{i=1}^{n-1}\equiv 0$, and a filter variable $s_n(x,t):\mathbb{R}^n\times[0,T)\rightarrow[0,T)$, where the $s_i,~i=2,\cdots,n$ are defined inductively by
	\begin{equation}\label{s}
	\begin{aligned}
	&s_1 = x_1, ~   \\
	&s_i  = \frac{k_{i-1}}{T-t} s_{i-1}+\dot{s}_{i-1} 
	\end{aligned}
	\end{equation}
	with $x_1$ being the output and $\{k_i\}_{i=1}^{n-1}$ satisfying $k_i>n-i+1$. If $s_n\rightarrow0$ as $t\rightarrow T$, then $\{s_{i}\}_{i=1}^{n-1}$ converges to zero as  $t\rightarrow T$.
\end{lemma}
\textit{Proof:} see Appendix II.  $\hfill\blacksquare $

\begin{remark}
	According to Lemma \ref{lemma-s}, one can define $$s_2=\frac{k_1}{T-t}x_1+x_2$$ for a second-order system, and define $$s_3=\frac{k_1k_2+k_1}{(T-t)^2}x_1+\frac{k_2+k_1}{T-t}x_2+x_3$$ for a third-order system, and so on.
	It can be seen that the filter variable $s_n(x,t)$ as defined  in Lemma \ref{lemma-s} is essentially different from the commonly used way of defining the filtered variable as $s_n=l_1x_1+l_2x_2+\cdots+x_n$ since a time-varying term $1/(T-t)$ is injected into $s_n$. Such treatment, together with other design skills, makes it possible to address  the adaptive prescribed-time control of the systems in normal-form (i.e., $x_1^{(n)}=b(t)u+\phi_n^{\top}\theta(t)$).
\end{remark}

\begin{lemma}[{[\citen{Huang2018,ye-2022-ct}]}]\label{lemma2}
	Given any positive smooth function $\sigma(t):[0,+\infty)\rightarrow \mathbb{R}^+$, the following inequality holds
	\begin{equation}
	|s|-\frac{s^2}{\sqrt{s^2+\sigma^2}}<\sigma, ~\forall s\in \mathbb{R}.
	\end{equation}
\end{lemma}

\section{Prescribed-time Control Design}\label{Control Design} 
\begin{figure*}[!]
	\begin{center}
		\includegraphics[height=8cm]{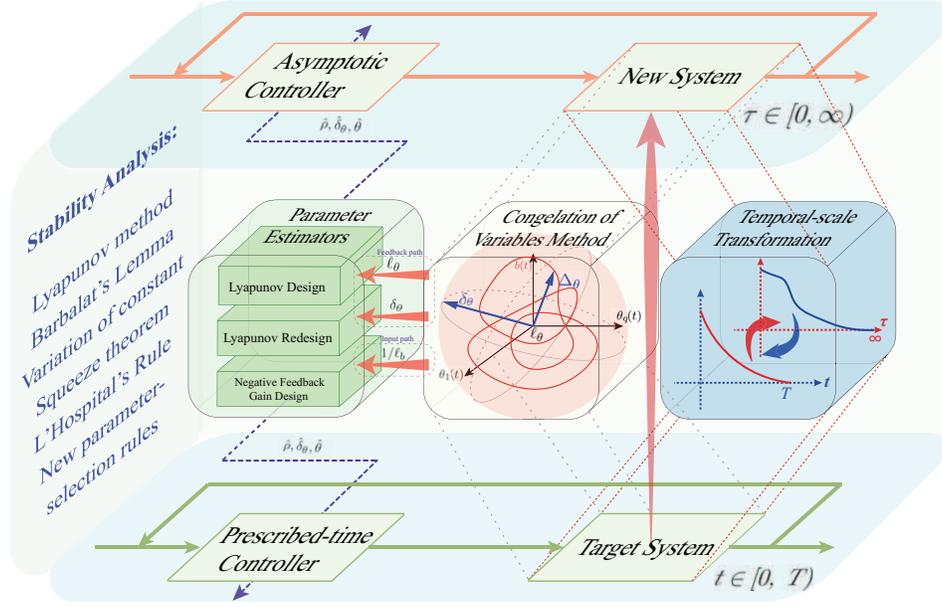}
		\caption{The design and analysis process and procedure  of the spatiotemporal transformation based adaptive prescribed-time control, where $\delta_{{\theta}}$ and $\ell_b$ are  defined in Assumptions \ref{assumption1}-\ref{assumption2}, $\ell_{\theta}$ can be regarded as the ``average" of $\theta(t)$,  $\Delta_{\theta}$ is a time-varying perturbation due to the inconsistency between $\theta(t)$ and $\ell_{\theta}$, and $\hat{\rho}$, $\hat{\delta}_\theta$ and $\hat{\theta}$ are the estimations of $1/\ell_{b}$, $\delta_{\theta}$, and $\ell_\theta$, respectively. We call the transformation of recasting the target system into a new system by temporal-scale transformation, generalized coordinate transformation and \textit{congelation of variable} method a spatiotemporal transformation.}\label{fig1}
	\end{center}
\end{figure*}

This section is devoted to establishing an adaptive prescribed-time control scheme of global stabilization for system (\ref{system}), which is nontrivial and demands several design techniques and transforms as described in Fig. \ref{fig1}. In particular,  three adaptive units and two robust units are incorporated, with the first adaptive unit estimating the ``average" of $\theta(t)$, the second estimating the ``radius" of $\Theta_0$, and the third estimating $1/\ell_{b}$,  while the  the first robust unit compensating the time-varying perturbations $\Delta_{\theta}$, and the second eliminating the lumped nonlinearities arisen from  the nonlinear damping design and the tuning functions design. The detailed control algorithms are  analyzed in the following subsections.


\subsection{System Reparameterization}\label{System Reparameterization}
It is interesting to note that by using the following temporal-scale transformation $\mathcal{T}_p:[0,T)\rightarrow[0,\infty)$ (inspired by \cite{2018-yucelen-finite,2020-yucelen-time-scaling,2020-NYU-dynamic-high-gain}) 
\begin{equation}
t =    T(1-e^{-\tau})  ~~\Leftrightarrow~~ \tau= \ln T-\ln (T-t) ,
\end{equation} 
where $T>0$ denotes the prescribed convergence time, we can transfer the interval $[0, T )$ in terms of the time variable $t$   to the interval  $[0,\infty)$ in terms of the time variable $\tau$. 
Thus for a signal $x(t)$ in $t$-axis, we can express it in $\tau$-axis such that $x(t)\equiv \breve x(\tau)$. Consequently, the dynamics $\dot x(t)=f(x,t)$ in $t$-axis can be equivalently described by 
\begin{equation}\label{beta}
\frac{d\breve x(\tau)}{d \tau}  =\frac{d t}{d \tau}\dot{x}(t) =\frac{d t}{d \tau} f( x,t)=\frac{d t}{d \tau} \breve f(\breve x,\tau) \triangleq \frac{1}{\beta(\tau) } \breve{f}(\breve{x},\tau)
\end{equation}
where $\beta(\tau)=e^{\tau}/T$ and its expression on $t$-axis is $\mu(t)=1/(T-t)$.

To proceed, we  perform  the general coordinate transformation as
\begin{equation}\label{error variables}
\begin{aligned} 
\alpha_0 =0,~ 
z_i  =  x_i-\alpha_{i-1}(\underline{x}_{i-1},\hat{\theta},\hat{\delta}_\theta,t), ~i=1,\cdots,n,~~
\end{aligned}
\end{equation}
where $\{\alpha_i(\cdot)\}_{i=1}^{n-1}$ are some smooth functions to be specified later. In addition, we define the new vectors as
\begin{equation}\label{new regressor vectors}
w_{i}\big(\underline{x}_{i}, \hat{\theta},\hat{\delta}_\theta,t\big) =\phi_{i}-\sum_{j=1}^{i-1} \frac{\partial \alpha_{i-1}}{\partial x_{j}} \phi_{j}.
\end{equation}
Furthermore, due to the presence of unknown time-varying parameters, we  use the  method of  \textit{congelation of variables}  to deal with these parameters, thereby   extracting the unknown constant parameters that can be used for certainty equivalence controller design. Namely,
\begin{equation}\label{10}
\begin{aligned}
\theta(t)&=\hat{\theta}+\big(\ell_{\theta}-\hat{\theta}\big)+\Delta_{\theta},\\
b(t)u&=\bar{u}- {\ell_{b}}\left(\frac{1}{\ell_{b}}-\hat{\rho}\right)\bar{u}+\hat{\rho}\bar{u}\Delta_{b},
\end{aligned}
\end{equation} 
where $\ell_{\theta}$ can be regarded as the ``average" of $\theta(t)$\cite{chenkaiwen}, which is not necessarily known, $\ell_{b}$ is defined in Assumption \ref{assumption2}, and $\Delta_{\theta}=\theta(t)-\ell_{\theta}$ and $\Delta_{b}=b(t)-\ell_{b} $ are unknown time-varying perturbation terms.  In addition, $\hat{\theta}$ is an ``estimate" of $\ell_\theta$, $\hat\rho$ is an ``estimate" of $1/\ell_b$, and $u=\hat{\rho}\bar{u}$.

Based upon the above treatments, we use the spatiotemporal transformation (as shown in Fig. \ref{fig1}) to  recast the original system  (which is well-defined on $[0,T)$) into a new form (which is well-defined on $[0,\infty)$), allowing us to address asymptotic stability for the new system instead of the prescribed-time stability for the original system.  Under such setting, we examine the dynamic model of the new system 
\begin{equation} \label{open-loop}
\small{\begin{aligned}
	\frac{ d \breve{z}_1}{d \tau}=&\frac{1}{\beta(\tau)}\left(\breve{\alpha}_1+\breve{z}_2+\breve w_1^{\top} \hat\theta+\breve w_1^{\top}\Delta_{\theta}\right)+\frac{1}{\beta(\tau)}\breve{w}_1^{\top}(\ell_{\theta}-\hat{\theta})\\
	\frac{ d \breve{z}_i}{d \tau}=&\frac{1}{\beta(\tau)}\left(\breve\alpha_i +f_{o_i}+\breve z_{i+1}+\breve w_i^{\top} \hat\theta+\breve w_i^{\top}\Delta_{\theta}\right)\\
	&+\frac{1}{\beta(\tau)} \breve w_i^{\top}(\ell_{\theta}-\hat{\theta})-\frac{\partial \breve\alpha_{i-1}}{\partial \hat{\theta}}\frac{\partial \hat{\theta}}{\partial \tau}-\frac{\partial \breve\alpha_{i-1}}{\partial \hat{\delta}_\theta}\frac{\partial \hat{\delta}_\theta}{\partial \tau} , \\
	\frac{ d \breve{z}_n}{d \tau}=&\frac{1}{\beta(\tau)}\left(\breve {\bar u}+f_{o_n}+\breve w_n^{\top} \hat\theta +\breve w_n^{\top}\Delta_{\theta}\right) \\
	& + \frac{1}{\beta(\tau)} \breve w_n^{\top}(\ell_{\theta}-\hat{\theta})   -\frac{\partial \breve\alpha_{n-1}}{\partial \hat{\theta}}\frac{\partial \hat{\theta}}{\partial \tau}-\frac{\partial \breve\alpha_{n-1}}{\partial \hat{\delta}_\theta}\frac{\partial \hat{\delta}_\theta}{\partial \tau}\\
	&-\frac{\ell_{b}}{\beta(\tau)}\left(\frac{1}{\ell_{b}}-\hat{\rho}\right)\breve{\bar{u}}+\frac{1}{\beta(\tau)}\hat{\rho}\breve{\bar{u}}\Delta_{b},
	\end{aligned}} 
\end{equation}
where, for $i=2,\cdots,n$,    
\begin{equation} \label{foi}
\begin{aligned}
f_{o_i}=  -\sum_{j=1}^{i-1}\frac{\partial\breve{\alpha}_{i-1}}{\partial\breve{x}_j}\breve{x}_{j+1}-\beta(\tau)\frac{\partial \breve\alpha_{i-1}}{\partial \beta(\tau)}\frac{\partial \beta(\tau)}{\partial \tau}.
\end{aligned} 
\end{equation}

\subsection{Lyapunov Design By means of Congealed Variables}\label{Lyapunov Design via Adaptive Backstepping}
In our technical development, one of the main obstacles is dealing with time-varying parameters $\theta(t)$ in order to avoid $\dot{\theta}(t)$ (sometimes even $\dot{\theta}(t)$ does not exist) appearing in the control design or stability analysis. To circumvent this obstacle, we propose a two-level estimation for $\theta(t)$. Specifically speaking, 
we first introduce a congealed variable $\ell_{\theta}$ to replace $\theta(t)$ and  then introduce another congealed variable $\delta_{{\theta}}$ to cope with the unknown time-varying perturbations caused by $\theta(t)-\ell_{\theta}$. Thereafter, in the Lyapunov design, it is sufficient to design the corresponding adaptive laws only for the congealed (time-invariant) parameters $\ell_{\theta}$ and $\delta_{{\theta}}$. The detailed steps are as follows.

\textit{Step $1$:} Choose a Lyapunov function defined on $[0,\infty)$  as 
\begin{equation}\label{V_theta}
V_{1\tau}=\frac{1}{2}\breve{z}_1^2+V_{\theta},
\end{equation}
where  $V_\theta=\frac{1}{2}(\ell_{\theta}-\hat\theta)^{\top}\Gamma_1^{-1}(\ell_{\theta}-\hat\theta)$ with $\Gamma_1\succ 0$.  Then,
\begin{equation} \label{dV1tau}
\begin{aligned}
\frac{d V_{1\tau}}{d\tau}=& \frac{1}{\beta(\tau)}\breve{z}_1\left(\breve{\alpha}_1+\breve{z}_2+\breve w_1^{\top} \hat\theta+\breve w_1^{\top} \Delta_\theta\right)\\
&+(\ell_{\theta}-\hat\theta)^{\top}\Gamma_1^{-1}\left( \frac{\Gamma_1}{\beta(\tau)}\breve{w}_1\breve{z}_1-\frac{d\hat\theta}{d\tau}\right).
\end{aligned}
\end{equation}
Design the virtual control law $\breve{\alpha}_1$ and the tuning function $\breve \tau_{\theta_1}$ as
\begin{equation} \label{alpha1}
\begin{aligned}
\breve{\alpha}_1 =-k_1\beta (\tau)\breve{z}_1-\breve{w}_1^{\top}\hat{\theta}+v_1(\breve z_1,\hat{\delta}_{\theta}) ,
\end{aligned}
\end{equation}
\begin{equation} \label{tau-theta-1}
\begin{aligned} 
\breve\tau_{\theta_1} = \breve{w}_1\breve{z}_1,
\end{aligned}
\end{equation}
where $k_1>n$, and $v_1$  will be designed in Section  \ref{Lyapunov Redesign with Tuning Functions}. Inserting (\ref{alpha1}) and (\ref{tau-theta-1}) into (\ref{dV1tau}), yields
\begin{equation} 
\begin{aligned}
\frac{d V_{1\tau}}{d\tau}=&-k_1\breve{z}_1^2+\frac{1}{\beta(\tau)}\left(\breve{z}_1v_1+\breve{z}_1\breve w_1^{\top}\Delta_\theta\right)+\frac{1}{\beta(\tau)}\breve{z}_{1}\breve{z}_{2}\\
&+(\ell_{\theta}-\hat\theta)^{\top}\Gamma_1^{-1}\left( \frac{\Gamma_1}{\beta(\tau)}\breve\tau_{\theta_1}-\frac{d\hat\theta}{d\tau}\right). 
\end{aligned}
\end{equation}

\textit{Step $2$:} Choose a Lyapunov function defined on $[0,\infty)$  as 
$V_{2\tau}=V_{1\tau} +\frac{1}{2}\breve{z}_2^2$, its derivative \textit{w.r.t.} $\tau$ is
\begin{equation} \label{dV2tau}
\small{ \begin{aligned}
	\frac{d V_{2\tau}}{d\tau}=&\frac{d V_{1\tau}}{d\tau}+ \frac{1}{\beta(\tau)}\breve{z}_2\left(\breve{\alpha}_2+f_{o2}+\breve{z}_3+\breve w_2^{\top} \hat\theta+\breve{w}_2^{\top}\Delta_{\theta}\right)  \\ 
	&+\frac{1}{\beta(\tau)}\breve{z}_2\breve{w}_2^{\top} (\ell_{\theta}-\hat{\theta}) -\breve{z}_2\left(\frac{\partial \breve\alpha_1}{\partial \hat\theta}\frac{\partial \hat\theta}{\partial \tau}+\frac{\partial \breve\alpha_1}{\partial \hat{\delta}_{\theta}}\frac{\partial \hat{\delta}_{\theta}}{\partial \tau}\right).
	\end{aligned}}
\end{equation}
Design the virtual control law $\breve{\alpha}_2$ and the tuning function $ \breve\tau_{\theta_2}$ as
\begin{equation} \label{alpha2}
\small{\begin{aligned}
	\breve{\alpha}_2 = -k_2\beta (\tau)\breve{z}_2-\breve{w}_2^{\top}\hat{\theta}+v_2(\underline{\breve z}_2,\hat{\delta}_{\theta})-f_{o2}+\frac{\partial \breve\alpha_1}{\partial \hat\theta}\Gamma_1\breve\tau_{\theta_2}-\breve{z}_1, 
	\end{aligned}}
\end{equation}
\begin{equation} \label{tau-theta-2}
\begin{aligned} 
\breve\tau_{\theta_2} = \tau_{\theta_1}+ \breve{w}_2\breve{z}_2,
\end{aligned}
\end{equation}
where $k_2>n-1$. Then, (\ref{dV2tau}) becomes
\begin{equation} 
\begin{aligned}
\frac{d V_{2\tau}}{d\tau}=& -\sum_{i=1}^{2}k_i\breve{z}_i^2+\frac{1}{\beta(\tau)}\sum_{i=1}^{2}\left(\breve{z}_iv_i+\breve{z}_i\breve{w}_i^{\top}\Delta_\theta\right)\\
&+\left(\frac{\partial \breve\alpha_1}{\partial\hat{\theta}}\breve{z}_2+(\ell_{\theta}-\hat{\theta})^{\top}\Gamma_1^{-1}\right)\left(\frac{\Gamma_1}{\beta(\tau)}\breve\tau_{\theta_2}-\frac{d\hat{\theta}}{d\tau}\right) \\
&+\frac{1}{\beta(\tau)}\breve{z}_2\breve{z}_3-\breve{z}_2\frac{\partial \breve{\alpha}_1}{\partial \hat{\delta}_{\theta}}\frac{\partial \hat{\delta}_{\theta}}{\partial \tau}.
\end{aligned}
\end{equation}

\textit{Step $i ~(i=3,\cdots,n-1)$:} Choose a Lyapunov function defined on $[0,\infty)$  as 
$V_{i\tau}=V_{{({i-1})}\tau} +\frac{1}{2}\breve{z}_i^2$. Then, the derivative of $V_{i\tau}$ along the trajectory of (\ref{open-loop}) is evaluated as
\begin{equation} \label{dVitau}
\small{\begin{aligned}
	\frac{d V_{i\tau}}{d\tau}=&\frac{d V_{(i-1)\tau}}{d\tau}+ \frac{1}{\beta(\tau)}\breve{z}_i\left(\breve{\alpha}_i+f_{oi}+\breve{z}_{i+1}+\breve w_i^{\top} \hat\theta+\breve{w}_i^{\top}\Delta_{\theta}\right)  \\ 
	&+\frac{1}{\beta(\tau)}\breve{z}_i\breve{w}_i^{\top} (\ell_{\theta}-\hat{\theta}) -\breve{z}_i\left(\frac{\partial \breve\alpha_{i-1}}{\partial \hat\theta}\frac{\partial \hat\theta}{\partial \tau}+\frac{\partial \breve\alpha_{i-1}}{\partial \hat{\delta}_{\theta}}\frac{\partial \hat{\delta}_{\theta}}{\partial \tau}\right).
	\end{aligned}}
\end{equation}
Design the virtual control law $\breve{\alpha}_i$ and the tuning function $ \breve\tau_{\theta_i}$ as
\begin{equation} \label{alpha-i}
\small{\begin{aligned}
	\breve{\alpha}_i=& -k_i\beta (\tau)\breve{z}_i-\breve{w}_i^{\top}\hat{\theta}+v_i(\underline{\breve z}_i,\hat{\delta}_{\theta})-f_{oi}+\frac{\partial \breve\alpha_{i-1}}{\partial \hat\theta}\Gamma_1\breve\tau_{\theta_i}\\
	&+\sum_{j=2}^{i-1}\frac{\partial \breve\alpha_{j-1}}{\partial \hat\theta}{\Gamma}_1\breve{w}_i\breve{z}_j-\breve{z}_{i-1},
	\end{aligned}}
\end{equation}
\begin{equation} \label{tau_i}
\breve\tau_{\theta_i}=  \breve\tau_{\theta_{i-1}}+ \breve{w}_i\breve{z}_i, 
\end{equation}
where $k_i>n-i+1$, and $v_i$  will be designed in Section  \ref{Lyapunov Redesign with Tuning Functions}. The resulting ${d V_{i\tau}}/{d\tau}$ is
\begin{equation} 
\small{\begin{aligned}
	\frac{d V_{i\tau}}{d\tau}=& -\sum_{j=1}^{i}k_j\breve{z}_j^2+\frac{1}{\beta(\tau)}\sum_{j=1}^{i}\left(\breve{z}_jv_j+\breve{z}_j\breve{w}_j^{\top}\Delta_\theta\right)\\
	&+\left(\sum_{j=1}^{i-1}\frac{\partial \breve\alpha_j}{\partial \hat\theta}\breve z_{j+1}+(\ell_{\theta}-\hat{\theta})^{\top}\Gamma_1^{-1}\right)\left(\frac{\Gamma_1}{\beta(\tau)}\breve\tau_{\theta_i}-\frac{d\hat{\theta}}{d\tau}\right) \\
	&+\frac{1}{\beta(\tau)}\breve{z}_{i}\breve{z}_{i+1}-\breve{z}_i\frac{\partial \breve{\alpha}_{i-1}}{\partial \hat{\delta}_{\theta}}\frac{\partial \hat{\delta}_{\theta}}{\partial \tau}.
	\end{aligned}}
\end{equation}

\textit{Step $n$:} Consider the Lyapunov function candidate as 
\begin{equation}\label{V_rho}
V_{n\tau}=V_{(n-1)\tau}+\frac{1}{2}\breve{z}_n^2+V_\rho,
\end{equation}
where $V_{\rho}=\frac{|\ell_{b}|}{2\gamma_{\rho}}\left(\frac{1}{\ell_{b}}-\hat{\rho}\right)^2$. Then,
\begin{equation} \label{dVntau}
\begin{aligned}
\frac{d V_{n\tau}}{d\tau}=&\frac{d V_{(n-1)\tau}}{d\tau}+ \frac{1}{\beta(\tau)}\breve{z}_n\left(\breve{\bar u}+f_{on}+ \breve w_n^{\top} \hat\theta+\breve{w}_n^{\top}\Delta_{\theta}\right)  \\ 
&+\frac{1}{\beta(\tau)}\breve{z}_n\breve{w}_n^{\top} (\ell_{\theta}-\hat{\theta})+\frac{1}{\beta(\tau)}\breve{z}_n\hat\rho\breve{\bar {u}}\Delta_{b}\\
& -\breve{z}_n\left(\frac{\partial \breve\alpha_{n-1}}{\partial \hat\theta}\frac{\partial \hat\theta}{\partial \tau}+\frac{\partial \breve\alpha_{n-1}}{\partial \hat{\delta}_{\theta}}\frac{\partial \hat{\delta}_{\theta}}{\partial \tau}\right)\\
&-  \frac{|\ell_{b}|}{\gamma_{\rho}}\left(\frac{1}{\ell_{b}}-\hat{\rho}\right)\left(\frac{\gamma_{\rho}}{\beta(\tau)}\operatorname{sgn}(\ell_{b})\breve{z}_n\breve{\bar u}+ \frac{d{\hat{\rho}}}{d\tau}\right).
\end{aligned}
\end{equation}
Design  the control input and update laws of $\hat{\theta}$ and $\hat{\rho}$ as follows:
\begin{equation} \label{bar_u}
\begin{aligned}
\breve{\bar u}=-k_n\beta (\tau)\breve{z}_n + \kappa(\underline{\breve x} ,\hat{\theta},\hat{\delta}_\theta,\tau)\breve z_n+v_n,
\end{aligned}
\end{equation}
\begin{equation} \label{d_theta}
\begin{aligned}
\frac{d\hat{\theta}}{d\tau}=\frac{\Gamma_1}{\beta(\tau)}\breve\tau_{\theta_n}=\frac{\Gamma_1}{\beta(\tau)}\sum_{i=1}^{n}\breve{w}_i\breve{z}_i,
\end{aligned}
\end{equation}
\begin{equation}\label{d_rho} 
\begin{aligned}
\frac{d {\hat{\rho}}}{d \tau}=-\frac{\gamma_{\rho}}{\beta(\tau)}\operatorname{sgn}(\ell_{b})\breve{z}_n\breve{\bar u},
\end{aligned}
\end{equation}
where $k_n>1/(\underline{b}\hat{\rho}(0))$, and   functions $v_n$ and $ \kappa(\cdot)>0$ will be designed in Sections \ref{Lyapunov Redesign with Tuning Functions} and \ref{Negative Feedback Gain Design}, respectively.  Then, (\ref{dVntau}) can be continued as follows,
\begin{equation} \label{dVntau_mid}
\begin{aligned}
\frac{d V_{n\tau}}{d\tau}
=& -\sum_{i=1}^{n}k_i\breve{z}_i^2   +\frac{1}{\beta(\tau)}\breve{z}_n\left(\breve{z}_{n-1}+\breve w_n^{\top} \hat\theta  +  \kappa(\cdot)\breve{z}_n \right.\\
&\left.+f_{on}-\sum_{j=2}^{n-1}\Gamma_1\frac{\partial \breve{\alpha}_{j-1}}{\partial \hat{\theta}}\breve{z}_j\breve{w}_n -\frac{\partial\breve \alpha_{n-1}}{\partial \hat{\theta}}\Gamma_1\breve\tau_{\theta_n}  \right) \\
& -\sum_{i=2}^{n}\breve{z}_{i}\frac{\partial \breve\alpha_{i-1}}{\partial \hat{\delta}_{\theta}}\frac{\partial \hat{\delta}_{\theta}}{\partial \tau} +\frac{1}{\beta(\tau)}\breve{z}_n\breve{u}\hat\rho\Delta_{b}   \\&
+\frac{1}{\beta(\tau)}\sum_{i=1}^{n}\left(\breve{z}_iv_i+\breve{z}_i\breve{w}_i^{\top}\Delta_\theta\right).
\end{aligned}
\end{equation}
\begin{remark}
	In the absence of time-varying parameters, such a prescribed-time stabilization problem was solved and well-understood when $b(t)=1$
	in (\ref{system}), as shown in \cite{2021-TAC-huachanghun-prescribed-adaptive}.  Its solution employs a non-state/temporal scaling design to achieve prescribed-time control for a class of strict-feedback systems with unknown time-invariant parameters in the feedback path. Note that if we choose $v_1,v_2,\cdots,v_i\equiv 0$ in (\ref{alpha1}), (\ref{alpha2}), and (\ref{alpha-i}), and design 
	the control input $\breve{\bar{u}}$  as
	\begin{equation}
	\begin{aligned}
	\breve{\bar u}=&-k_n\beta (\tau)\breve{z}_n-\breve{w}_n^{\top}\hat{\theta} -f_{on}+\frac{\partial \breve\alpha_{n-1}}{\partial \hat\theta}\Gamma_1\breve\tau_{\theta_i}\\
	&+\sum_{j=2}^{n-1}\frac{\partial \breve\alpha_{j-1}}{\partial \hat\theta}\Gamma_1\breve{w}_n\breve{z}_j-\breve{z}_{n-1},~~k_n>n,
	\end{aligned}
	\end{equation}
	then such control algorithm is reduced to the one proposed in [\citen{2021-TAC-huachanghun-prescribed-adaptive}, Theorem 2].
	For issues in adaptive prescribed-time control for parameter-varying systems (e.g., estimation of time-varying parameters, lumped negative feedback design, stability analysis, boundedness analysis of control input and update laws, etc.), new solutions are
	needed, which will be developed in the next sections.
\end{remark}

\subsection{Lyapunov Redesign With Tuning Functions}\label{Lyapunov Redesign with Tuning Functions}
Aiming at the last line of (\ref{dVntau_mid}), performing certainty equivalence principle on $v_i$ and deliberately adding nonlinear damping terms, we redesign a part of the virtual/actual control inputs $v_i$ step by step   to  offset $\frac{1}{\beta}\breve{z}_i\breve{w}_i^{\top}\Delta_{\theta}$ while canceling $-\sum_{i=2}^{n}\breve{z}_{i}\frac{\partial \breve\alpha_{i-1}}{\partial \hat{\delta}_{\theta}}\frac{\partial \hat{\delta}_{\theta}}{\partial \tau}$. 
Applying Lemma \ref{lemma2}, we have
\begin{equation}\label{33}
\small{\begin{aligned}
	\frac{1}{\beta}\breve{z}_i\breve{w}_i^{\top}\Delta_{\theta}&\leq \frac{1}{\beta}\delta_{{\theta}}|\breve{z}_i|\sqrt{\breve{w}_i^{\top}\breve{w}_i}\leq  \frac{\delta_{{\theta}} \varepsilon}{\beta} +\frac{\delta_{{\theta}}\breve{z}_i^2 \breve{w}_i^{\top}\breve{w}_i}{\beta\sqrt{\breve{z}_i^2\breve{w}_i^{\top}\breve{w}_i+ \varepsilon^2}},
	\end{aligned}}
\end{equation}
where $\gamma_{\delta}>0$, $\delta_{\theta}=\sqrt{\Delta_{\theta}^\top\Delta_{\theta}}$, $ \varepsilon>0$ is a constant.
\begin{remark}
	In \cite{chenkaiwenIFAC,chenkaiwen,yehefu}, it is necessary to find a regression vector matrix $W_i$ such that $w_i$ can be decomposed as $w_i=W_i\underline{z}_i$. This task is in fact not easy and its complexity increases significantly with the order of the system. In the present work, we give an alternative scheme (e.g. Eq. (\ref{33})) that successfully circumvents this technical obstacle, making the algorithm computationally simpler and easier to implement. Note that we do not need to design additional feedback terms to cancel ${\delta_{{\theta}} \varepsilon}/{\beta}$ in (\ref{33}), as this term will naturally converge to zero as $\tau\rightarrow\infty$,  thus not affecting the asymptotic convergence properties of the closed-loop system on $[0,\tau)$.
\end{remark}

\textit{Step 1:} We first choose a Lyapunov function as     
\begin{equation}\label{V_delta}
V_{\delta}=\frac{1}{2\gamma_\delta}\left(\delta_{{\theta}}-\hat{\delta}_\theta\right)^2=\frac{1}{2\gamma_\delta} \tilde{\delta}_\theta^2
\end{equation}  
with $\tilde{\delta}_\theta=\delta_\theta-\hat{\delta}_\theta$. Design 
\begin{equation}
v_1=-  \frac{{\hat{\delta}_\theta}  \breve{z}_1  \breve{w}_1^{\top}\breve{w}_1}{\sqrt{\breve{z}_1^2\breve{w}_1^{\top}\breve{w}_1+ \varepsilon^2}} ,~~
\breve\tau_{\delta_1}= \frac{   \breve{z}_1 ^2 \breve{w}_1^{\top}\breve{w}_1}{\sqrt{\breve{z}_1^2\breve{w}_1^{\top}\breve{w}_1+ \varepsilon^2}}  ,
\end{equation}
such that
\begin{equation*} \label{z1v1}
\begin{aligned}
\frac{1}{\beta }\left(\breve z_1v_1+\breve{z}_1\breve{w}_1^{\top}\Delta_{\theta}\right) +\frac{d V_{\delta}}{d \tau}\leq       \frac{\tilde{\delta}_\theta}{ \gamma_{\delta}}  \left(\frac{\gamma_{\delta} }{\beta }\breve\tau_{\delta_{1}} -\frac{d\hat{\delta}_\theta}{d\tau}\right)+ \frac{\delta_{{\theta}} \varepsilon }{\beta} .
\end{aligned}
\end{equation*}
Meanwhile, the virtual control $\breve\alpha_1$ can be rewritten as  
\begin{equation}
\breve{\alpha}_1=-k_1\beta\breve{z}_1-\breve{w}_1^\top\hat{\theta}- \frac{{\hat{\delta}_\theta}  \breve{z}_1  \breve{w}_1^{\top}\breve{w}_1}{\sqrt{\breve{z}_1^2\breve{w}_1^{\top}\breve{w}_1+ \varepsilon^2}} .
\end{equation}  
Note that it is easy to prove $\hat{\delta}_\theta>0$ by calling (\ref{update law of delta}). Thus we can conclude that $k_1=\inf\{\mathsf{K}_1(\tau)\}$ according to the expression of $\mathsf{K}_1$ as shown in (\ref{compact-set}). This conclusion, together with Lemma \ref{lemma1} and the subsequent design, allows for a rigorous proof of the boundedness of $\breve\alpha_1$, as seen in Section \ref{stability-analysis}.

\textit{Step 2:}     Design
\begin{equation}
v_2=-\frac{{\hat{\delta}_\theta}  \breve{z}_2 \breve{w}_2^{\top}\breve{w}_2}{\sqrt{\breve{z}_2^2\breve{w}_2^{\top}\breve{w}_2+ \varepsilon^2}} +\gamma_{\delta}\frac{\partial \breve\alpha_{1}}{\partial\hat\delta_{{\theta}}}\breve{\tau}_{\delta_2} ,
\end{equation}
with
$
\breve\tau_{\delta_2}=\breve\tau_{\delta_{1}}+\frac{   \breve{z}_2^2 \breve{w}_2^{\top}\breve{w}_2}{\sqrt{\breve{z}_2^2\breve{w}_2^{\top}\breve{w}_2+ \varepsilon^2}},
$
such that
\begin{equation} \label{z2v2}
\begin{aligned}
&\sum_{i=1}^2\frac{1}{\beta }\left(\breve z_iv_i+\breve{z}_i\breve{w}_i^{\top}\Delta_{\theta}\right) +\frac{d V_{\delta}}{d \tau}
\\& ~~~~ \leq     \frac{\tilde{\delta}_\theta}{ \gamma_{\delta}}  \left(\frac{\gamma_{\delta} }{\beta }\breve\tau_{\delta_{2}} -\frac{d\hat{\delta}_\theta}{d\tau}\right)+ \frac{2\delta_{{\theta}} \varepsilon }{\beta}+ \frac{\gamma_{\delta}}{\beta}\frac{\partial \breve\alpha_{1}}{\partial\hat\delta_{{\theta}}}\breve{z}_2\breve{\tau}_{\delta_2}.
\end{aligned}
\end{equation}
Hence, $\breve\alpha_2$ is designed as
\begin{equation} \label{alpha2-new}
\begin{aligned}
\breve{\alpha}_2 =& -k_2\beta \breve{z}_2-\breve{w}_2^{\top}\hat{\theta}
-\frac{{\hat{\delta}_\theta}  \breve{z}_2 \breve{w}_2^{\top}\breve{w}_2}{\sqrt{\breve{z}_2^2\breve{w}_2^{\top}\breve{w}_2+ \varepsilon^2 }}+\gamma_{\delta}\frac{\partial \breve\alpha_{1}}{\partial\hat\delta_{{\theta}}}\breve{\tau}_{\delta_2}\\
&  +\frac{\partial\breve{\alpha}_{1}}{\partial\breve{x}_1}\breve{x}_{2}+ \beta \frac{\partial \breve \alpha_1}{\partial \beta}\frac{\partial \beta}{\partial \tau}  +\frac{\partial \breve\alpha_1}{\partial \hat\theta}\Gamma_1\breve\tau_{\theta_2}-\breve{z}_1.
\end{aligned}
\end{equation} 

\textit{Step $i$ ($i=3,\cdots,n-1$):} Motivated by the above design skills, we  design  the nonlinear damping terms  
\begin{equation}\label{vi}
\begin{aligned}
v_i=&-\frac{{\hat{\delta}_\theta}  \breve{z}_i \breve{w}_i^{\top}\breve{w}_i}{\sqrt{\breve{z}_i^2\breve{w}_i^{\top}\breve{w}_i+ \varepsilon^2}}+\gamma_{\delta}\frac{\partial \breve\alpha_{i-1}}{\partial\hat\delta_{{\theta}}}\breve{\tau}_{\delta_i}\\
& + \frac{\gamma_{\delta}   \breve{z}_i  \breve{w}_i^{\top}\breve{w}_i}{\sqrt{\breve{z}_i^2\breve{w}_i^{\top}\breve{w}_i+ \varepsilon^2}}  \sum_{j=2}^{i-1}\frac{\partial \breve\alpha_{j-1}}{\partial \hat\delta_\theta}  \breve{z}_j,
\end{aligned}
\end{equation}
with  
$
\tau_{\delta_i}=\tau_{\delta_{i-1}}+\frac{   \breve{z}_i^2 \breve{w}_i^{\top}\breve{w}_i}{\sqrt{\breve{z}_i^2\breve{w}_i^{\top}\breve{w}_i+ \varepsilon^2}},
$
such that 
\begin{equation} \label{zivi}
\small{\begin{aligned}
	&\sum_{j=1}^i\frac{1}{\beta }\left(\breve z_jv_j+\breve{z}_j\breve{w}_j^{\top}\Delta_{\theta}\right) +\frac{d V_{\delta}}{d \tau}
	\\ 
	& ~~~~\leq \sum_{j=3}^i\frac{1}{\beta}\left(\breve{z}_jv_j+\delta_{{\theta}} \varepsilon  +\frac{\delta_{{\theta}}\breve{z}_j^2 \breve{w}_j^{\top}\breve{w}_j}{\sqrt{\breve{z}_j^2\breve{w}_j^{\top}\breve{w}_j+ \varepsilon^2 }}\right)\\
	& 
	~~~~~~~+ \frac{\tilde{\delta}_\theta}{ \gamma_{\delta}}  \left(\frac{\gamma_{\delta} }{\beta }\breve\tau_{\delta_{2}} -\frac{d\hat{\delta}_\theta}{d\tau}\right)+ \frac{2\delta_{{\theta}} \varepsilon }{\beta} + \frac{\gamma_{\delta}}{\beta}\frac{\partial \breve\alpha_{1}}{\partial\hat\delta_{{\theta}}}\breve{z}_2\breve{\tau}_{\delta_2}\\
	\end{aligned}} 
\end{equation}
\begin{equation*}  
\begin{aligned}
&~~~~= \frac{\tilde{\delta}_\theta}{ \gamma_{\delta}}  \left(\frac{\gamma_{\delta} }{\beta }\breve\tau_{\delta_{i}} -\frac{d\hat{\delta}_\theta}{d\tau}\right)+ \frac{i \delta_{{\theta}} \varepsilon }{\beta} +\sum_{j=2}^{i} \frac{\gamma_{\delta}}{\beta}\frac{\partial \breve\alpha_{j-1}}{\partial\hat\delta_{{\theta}}}\breve{z}_j\breve{\tau}_{\delta_j}\\
&~~~~~~~+ \sum_{j=1}^{i}\frac{\gamma_{\delta}}{\beta}\left(\frac{  \breve{z}_j ^2 \breve{w}_j^{\top}\breve{w}_j}{\sqrt{\breve{z}_j^2\breve{w}_j^{\top}\breve{w}_j+ \varepsilon^2 }}  \sum_{p=2}^{j-1}\frac{\partial \breve\alpha_{p-1}}{\partial \hat\delta_\theta}  \breve{z}_p\right).
\end{aligned} 
\end{equation*} 

\textit{Step n:}  Slightly different from the above design steps, we design
\begin{equation} \label{vn}
v_n= -\frac{{\hat{\delta}_\theta}  \breve{z}_n \breve{w}_n^{\top}\breve{w}_n}{\sqrt{\breve{z}_n^2\breve{w}_n^{\top}\breve{w}_n+ \varepsilon^2 }} 
\end{equation} 
such that no positive feedback terms are included in the final control input $\hat{\rho}\breve{\bar{u}}$. To go on,  we design the update law of $\hat{\delta}_\theta$ as 
\begin{equation} \label{update law of delta}
\frac{d\hat{\delta}_\theta}{d \tau}=\frac{\gamma_{\delta}}{\beta}\breve\tau_{\delta_{n}} =\frac{\gamma_{\delta}}{\beta}\sum_{i=1}^n\frac{   \breve{z}_i^2 \breve{w}_i^{\top}\breve{w}_i}{\sqrt{\breve{z}_i^2\breve{w}_i^{\top}\breve{w}_i+ \varepsilon^2 }},
\end{equation} 
with
\begin{equation} \label{tau-delta_i} 
\breve\tau_{\delta_{n}}= \tau_{\delta_{n-1}}+\frac{   \breve{z}_n^2 \breve{w}_n^{\top}\breve{w}_n}{\sqrt{\breve{z}_n^2\breve{w}_n^{\top}\breve{w}_n+ \varepsilon^2 }}.
\end{equation}
Then, it holds that
\begin{equation} \label{znvn-new}
\begin{aligned}
&\sum_{i=1}^n\frac{1}{\beta }\left(\breve z_iv_i+\breve{z}_i\breve{w}_i^{\top}\Delta_{\theta}\right) +\frac{d V_{\delta}}{d \tau} -\sum_{i=2}^{n}\breve{z}_{i}\frac{\partial \breve\alpha_{i-1}}{\partial \hat{\delta}_{\theta}}\frac{\partial \hat{\delta}_{\theta}}{\partial \tau}
\\
& \leq  \frac{n\delta_{{\theta}} {\varepsilon}}{\beta}
- \frac{\gamma_{\delta}}{\beta}\frac{\partial \breve\alpha_{n-1}}{\partial\hat\delta_{{\theta}}}\breve{z}_n\breve{\tau}_{\delta_n}
- \frac{\gamma_{\delta}}{\beta}\frac{   \breve{z}_n^2  \breve{w}_n^{\top}\breve{w}_n}{\sqrt{\breve{z}_n^2\breve{w}_n^{\top}\breve{w}_n+ \varepsilon^2 }} \\
&~~~\times \sum_{j=2}^{n-1}\frac{\partial \breve\alpha_{j-1}}{\partial \hat\delta_\theta}  \breve{z}_j .
\end{aligned}
\end{equation}
Choose
\begin{equation}\label{47}
V=V_z+V_\theta+V_{\rho}+V_{\delta},
\end{equation}
where $V_z=\frac{1}{2}\sum_{i=1}^{n}\breve{z}_i^2$, $V_\theta,~V_{\rho}$ and $V_{\delta}$ are defined in (\ref{V_theta}), (\ref{V_rho}) and (\ref{V_delta}), respectively. Combining (\ref{dVntau_mid}) and (\ref{znvn-new}), it is not difficult to deduce that
\begin{equation}  \label{dV-dtau}
\begin{aligned}
\frac{d V }{d\tau}
\leq& -\sum_{i=1}^{n}k_i\breve{z}_i^2   +\frac{1}{\beta(\tau)}\breve{z}_n\left(\breve{z}_{n-1}+\breve w_n^{\top} \hat\theta  +  \kappa(\cdot)\breve{z}_n \right.\\
&\left.+f_{on}-\breve\Psi_{\theta} -\breve\Psi_{{\delta}_{\theta}}\right) +\frac{1}{\beta(\tau)}\breve{z}_n\breve{u}\hat\rho\Delta_{b} + \frac{n\delta_{{\theta}} {\varepsilon}}{\beta},
\end{aligned}
\end{equation}
where 
\begin{equation}\label{Psi_theta}
\breve\Psi_{\theta}= \frac{\partial\breve \alpha_{n-1}}{\partial \hat{\theta}}\Gamma_1\breve\tau_{\theta_n} +\sum_{j=2}^{n-1}\Gamma_1\frac{\partial \breve{\alpha}_{j-1}}{\partial \hat{\theta}}\breve{z}_j\breve{w}_n ,
\end{equation} 
\begin{equation} \label{Psi_delta_theta}
\breve\Psi_{\delta_{\theta}}= 
{\gamma_{\delta}} \frac{\partial \breve\alpha_{n-1}}{\partial\hat\delta_{{\theta}}} \breve{\tau}_{\delta_n} +  \frac{   {\gamma_{\delta}} \breve{z}_n   \breve{w}_n^{\top}\breve{w}_n}{\sqrt{\breve{z}_n^2\breve{w}_n^{\top}\breve{w}_n+\breve\varepsilon^2 }}  \sum_{j=2}^{n-1}\frac{\partial \breve\alpha_{j-1}}{\partial \hat\delta_\theta}  \breve{z}_j.
\end{equation}  
Therefore,  (\ref{dV-dtau}) can be  continued as follows,
\begin{equation}\label{dVn}
\begin{aligned}
\frac{d V }{d\tau}
\leq&  -\sum_{i=1}^{n}k_i\breve{z}_i^2  +\frac{1}{\beta(\tau)}\breve{z}_n\breve{u}\hat\rho\Delta_{b}+\frac{n\delta_{{\theta}} {\varepsilon}}{\beta(\tau)}\\
& +\frac{1}{\beta(\tau)}\breve{z}_n\left[   \kappa(\breve{x} ,\hat{\theta},\hat{\delta}_\theta,\tau)\breve{z}_n+  \breve\Psi( {\breve x},\hat{\theta},\hat{\delta}_\theta,\tau) \right] ,
\end{aligned}
\end{equation}  
where 
\begin{equation}\label{Psi}
\breve\Psi(\breve{x},\hat{\theta},\hat{\delta}_\theta,\tau)=\breve{z}_{n-1}+\breve w_n^{\top} \hat\theta  +     f_{on}  -\breve\Psi_{\theta}-\breve\Psi_{\delta_{\theta}}.
\end{equation} 
Note that the time-varying terms $\Delta_{\theta}$ and $\Delta_{b}$  are injected into  (\ref{open-loop}) to reflect the effects of time-varying parameters on the dynamics of uncertain systems. In this subsection,  through Lyapunov redesign, we develop a low conservative scheme to deal with $\Delta_\theta$ while making the algorithm robust to uncertainties. Thereafter, we will dispose of $\Delta_{b}$ by negative feedback gain design in the next subsection.

\subsection{Negative Feedback Gain Design}\label{Negative Feedback Gain Design}
Here we rewrite $\breve{\bar u}$ as
\begin{equation}\label{breve_u}
\breve{\bar u}=-k_n\beta \breve{z}_n -  \frac{{\hat{\delta}_\theta}  \breve{z}_n \breve{w}_n^{\top}\breve{w}_n}{\sqrt{\breve{z}_n^2\breve{w}_n^{\top}\breve{w}_n+ \varepsilon^2 }} + {\kappa}(\cdot)\breve z_n.
\end{equation}
Design 
\begin{equation}
\kappa(\breve{x},\hat{\theta},\hat{\delta}_\theta,\tau)=-\frac{\breve\Psi^2}{\sqrt{\breve{z}_n^2\breve\Psi^2+ {\varepsilon}^2 }} ,
\end{equation}
while applying Lemma \ref{lemma2}, such that
\begin{equation}
\begin{aligned}
\kappa(\cdot)\breve{z}_n^2+\breve{\Psi}\breve{z}_n&=-\frac{\breve{z}_n^2\breve\Psi^2}{\sqrt{\breve{z}_n^2\breve\Psi^2+ {\varepsilon}^2 }}+\breve{z}_n\breve{\Psi}\\
&\leq-|\breve{z}_n\breve{\Psi}|+\breve{z}_n\breve{\Psi}+ {\varepsilon} \leq  {\varepsilon} .
\end{aligned}
\end{equation}
Now,  one can find that the control gain is always negative and hence the input $\breve{\bar u}$ can be rewritten as   $\breve{\bar u}=-\mathbf{K} \breve{z}_n$, with
\begin{equation}\label{K}
\begin{aligned}
\mathbf{K} =&k_n\beta  +  \frac{{\hat{\delta}_\theta}  \breve{w}_n^{\top}\breve{w}_n}{\sqrt{\breve{z}_n^2\breve{w}_n^{\top}\breve{w}_n+ \varepsilon^2 }}  +\frac{\breve\Psi^2}{\sqrt{\breve{z}_n^2\breve\Psi^2+ {\varepsilon}^2 }}>0.
\end{aligned}
\end{equation}
The resulting ${d V }/{d\tau}$ is
\begin{equation}\label{dVn-1}
\begin{aligned}
\frac{d V }{d\tau}
\leq&  -\sum_{i=1}^{n}k_i\breve{z}_i^2+\frac{(n\delta_{{\theta}}+1) {\varepsilon}}{\beta(\tau)}   -\frac{\mathbf{K} }{\beta(\tau)}\breve{z}_n^2 \hat\rho\Delta_{b}.
\end{aligned}
\end{equation}  
To close this section, we analyze the dynamic behaviour of the last term  on the right-hand side of  (\ref{dVn-1}), applying (\ref{d_rho}), (\ref{breve_u}) and (\ref{K}), yields
\begin{equation} \label{shedong}
\begin{aligned}
-\frac{\mathbf{K} }{\beta(\tau)}\breve{z}_n^2 \hat\rho\Delta_{b} =&-\frac{\mathbf{K}\breve{z}_n^2}{\beta(\tau)}\int_{0}^{\tau} \frac{\mathbf{K}\gamma_{\rho}}{\beta(s)}\breve{z}_n^2 ds \times  \operatorname{sgn}(\ell_{b})\Delta_{b}\\
& -\frac{\mathbf{K}\breve{z}_n^2}{\beta(\tau)}\hat{\rho}(0)\Delta_{b},
\end{aligned}
\end{equation}
where $\operatorname{sgn}(\ell_{b})\Delta_{b}>0 $.
According to Assumption  \ref{assumption2}, the direction of control is known and it can be positive or negative, so we discuss both cases separately.
\begin{itemize}
	\item Case 1: when $b(t)>0$, we have $\Delta_{b}(t)\geq0$  and $0<\ell_{b} \leq b(t)$, which means that any initialization with $\hat{\rho}(0)>0$ guarantees that the second line of (\ref{shedong}) is non-positive. 
	\item Case 2: when $b(t)<0$, we have  $\Delta_{b}(t)\leq 0$ and $b(t)\leq \ell_{b} <0  $, thereby $-\frac{\mathbf{K}\breve{z}_n^2}{\beta(\tau)}\hat{\rho}(0)\Delta_{b}\leq0$  can be guaranteed by selecting $\hat{\rho}(0)<0$.
\end{itemize}
Note that both cases imply that $\operatorname{sgn}(\ell_{b})\Delta_{b}\geq 0$. In view of (\ref{K}), it follows from  $\operatorname{sgn}(\ell_{b})\Delta_{b}\geq 0$  that $-\frac{\mathbf{K}}{\beta(\tau)}\breve{z}_n^2 \hat\rho\Delta_{b}\leq 0$, and 
\begin{equation}\label{DVn}
\begin{aligned}
\frac{d V }{d\tau}
\leq&  -\sum_{i=1}^{n}k_i\breve{z}_i^2 +\frac{(n\delta_{{\theta}}+1) {\varepsilon} }{\beta},
\end{aligned}
\end{equation}  
where $n$ is the system order, $\delta_{{\theta}}$ and $\varepsilon$ are  unknown constants, and $\beta$ is a monotonically increasing function as defined below (\ref{beta}).

\section{Main Results \& Stability analysis}\label{stability-analysis}
\subsection{Main Theorems}
\begin{table*}[!htbp]
	\caption{The asymptotic controller  on time domain $[0,\infty)$ and the prescribed-time controller on time domain $[0,T)$}
	\centering 
	\begin{tabular}{|l|l|}
		\hline
		\textbf{Asymptotic Controller $\breve u(\tau)$ with $\tau\in[0,\infty)$:}& \textbf{Prescribed-time Controller $u(t)$ with $t\in[0,T)$:}\\ 
		$\breve\alpha_1(\tau)=-k_1\beta\breve{z}_1-\breve{w}_1^\top\hat{\theta}- \frac{{\hat{\delta}_\theta}  \breve{z}_1  \breve{w}_1^{\top}\breve{w}_1}{\sqrt{\breve{z}_1^2\breve{w}_1^{\top}\breve{w}_1+ \varepsilon^2}} $,  
		& 	
		$ \alpha_1(t)=-k_1\mu(t) {z}_1- {w}_1^\top\hat{\theta}- \frac{{\hat{\delta}_\theta} {z}_1   {w}_1^{\top} {w}_1}{\sqrt{ {z}_1^2 {w}_1^{\top} {w}_1+ \varepsilon^2}} ,$  
		\\ 
		$\breve\alpha_i(\tau)=  -k_i\beta \breve{z}_i-\breve{w}_i^{\top}\hat{\theta}
		+ \frac{\gamma_{\delta} \breve{z}_i  \breve{w}_i^{\top}\breve{w}_i}{\sqrt{\breve{z}_i^2\breve{w}_i^{\top}\breve{w}_i+ \varepsilon^2}}  \sum_{j=2}^{i-1}\frac{\partial \breve\alpha_{j-1}}{\partial \hat\delta_\theta}  \breve{z}_j	
		$
		& 	$  \alpha_i(t)=  -k_i\mu(t) {z}_i- {w}_i^{\top}\hat{\theta}
		+ \frac{\gamma_{\delta}  {z}_i   {w}_i^{\top} {w}_i}{\sqrt{ {z}_i^2 {w}_i^{\top} {w}_i+ \varepsilon^2}}  \sum_{j=2}^{i-1}\frac{\partial  \alpha_{j-1}}{\partial \hat\delta_\theta}   {z}_j$    
		\\ 
		$~-\frac{{\hat{\delta}_\theta}  \breve{z}_i \breve{w}_i^{\top}\breve{w}_i}{\sqrt{\breve{z}_i^2\breve{w}_i^{\top}\breve{w}_i+ \varepsilon^2}}
		+\gamma_{\delta}\frac{\partial \breve\alpha_{i-1}}{\partial\hat\delta_{{\theta}}}\breve{\tau}_{\delta_i}	 
		+\sum_{j=1}^{i-1}\frac{\partial\breve{\alpha}_{i-1}}{\partial\breve{x}_j}\breve{x}_{j+1}-\breve{z}_{i-1}$
		& $~-\frac{{\hat{\delta}_\theta}  {z}_i  {w}_i^{\top} {w}_i}{\sqrt{ {z}_i^2 {w}_i^{\top} {w}_i+ \varepsilon^2}}
		+\gamma_{\delta}\frac{\partial  \alpha_{i-1}}{\partial\hat\delta_{{\theta}}} {\tau}_{\delta_i}	 
		+\sum_{j=1}^{i-1}\frac{\partial {\alpha}_{i-1}}{\partial {x}_j} {x}_{j+1}- {z}_{i-1} $
		\\ 
		$~ +\sum_{j=2}^{i-1}\frac{\partial \breve\alpha_{j-1}}{\partial \hat\theta}\Gamma_1\breve{w}_i\breve{z}_j+\frac{\partial \breve\alpha_{i-1}}{\partial \hat\theta}\Gamma_1\breve\tau_{\theta_i} +\beta\frac{\partial \breve\alpha_{i-1}}{\partial \beta}\frac{\partial \beta}{\partial \tau},$
		& 
		$~+\sum_{j=2}^{i-1}\frac{\partial  \alpha_{j-1}}{\partial \hat\theta}\Gamma_1 {w}_i {z}_j+\frac{\partial  \alpha_{i-1}}{\partial \hat\theta}\Gamma_1 \tau_{\theta_i}+ \sum_{j=0}^{i-2}\frac{\partial \alpha_{i-1}}{\partial \mu^{(j)}}\frac{\partial \mu^{(j)}}{\partial t},$
		\\   
		$\breve{\bar u}(\tau)= -\Big[k_n\beta  +  \frac{{\hat{\delta}_\theta}  \breve{w}_n^{\top}\breve{w}_n}{\sqrt{\breve{z}_n^2\breve{w}_n^{\top}\breve{w}_n+ \varepsilon^2 }}  +\frac{\breve\Psi^2}{\sqrt{\breve{z}_n^2\breve\Psi^2+ {\varepsilon}^2 }}\Big]\breve z_n,$ 
		&$ {\bar u}(t)= -\Big[k_n\mu +  \frac{{\hat{\delta}_\theta}   {w}_n^{\top} {w}_n}{\sqrt{ {z}_n^2{w}_n^{\top} {w}_n+\ \varepsilon^2 }}  +\frac{ \Psi^2}{\sqrt{\breve{z}_n^2 \Psi^2+ {\varepsilon}^2 }}\Big] z_n	 $ 
		\\
		$\breve u(\tau)=\hat{\rho}(\tau)\breve{\bar u}(\tau),$
		&
		$  u(t)=\hat{\rho}(t) {\bar u}(t)$\\
		\hline 
		\textbf{Asymptotic Parameter  Estimators:}& \textbf{Prescribed-time Parameter Estimators:}
		\\$\frac{d\hat{\theta}}{d\tau}  =\frac{\Gamma_1}{\beta}\sum_{i=1}^{n}\breve{w}_i\breve{z}_i,~~~~~~~~~~~~~~~~~~~~~~~~~~~~~~\hat\theta(0)>0$ 
		&$\dot{\hat{\theta}}  = \Gamma_1\sum_{i=1}^{n} {w}_i {z}_i,~~~~~~~~~~~~~~~~~~~~~~~~~~~~~~~\hat\theta(0)>0$  
		\\$\frac{d\hat{\delta}_\theta}{d \tau}= \frac{\gamma_{\delta}}{\beta}\sum_{i=1}^n\frac{   \breve{z}_i^2 \breve{w}_i^{\top}\breve{w}_i}{\sqrt{\breve{z}_i^2\breve{w}_i^{\top}\breve{w}_i+ \varepsilon^2 }},~~~~~~~~~~~~~ ~~~~ \hat\delta_\theta(0)>0$	
		&$ {\dot{\hat{\delta}}_\theta} =  \sum_{i=1}^n\frac{   \breve{z}_i^2 \breve{w}_i^{\top}\breve{w}_i}{\sqrt{\breve{z}_i^2\breve{w}_i^{\top}\breve{w}_i+ \varepsilon^2 }},~~~~~~~~~~~~~~~~~ ~~~~ \hat\delta_\theta(0)>0$
		\\ 
		$\frac{d {\hat{\rho}}}{d \tau}= \frac{-\gamma_{\rho}}{\beta}\operatorname{sgn}(\ell_{b})\breve{z}_n\breve{\bar u},~\hat{\rho}(0)>0~ \text{for} ~b(\tau)>0~ \&~ \hat{\rho}(0)<0~ \text{for}  ~b(\tau)<0.$
		&$ {\dot {\hat{\rho}}}  =- {\gamma_{\rho}} \operatorname{sgn}(\ell_{b}) {z}_n {\bar u},   ~\hat{\rho}(0)>0~ \text{for} ~b(t)>0~ \&~ \hat{\rho}(0)<0~ \text{for}  ~b(t)<0.$
		\\ 
		\hline
	\end{tabular}\vspace{0.3em}
	\footnotesize {The explicit expressions of $\breve z_i,~\breve{w}_i,~\breve\tau_{\theta_i},~\breve\tau_{\delta_{i}}$ for $i=1,\cdots,n$ and $\breve{ \Psi}$  can be found in (\ref{error variables}), (\ref{new regressor vectors}), (\ref{tau_i}), (\ref{tau-delta_i}), (\ref{Psi_theta}), (\ref{Psi_delta_theta}), and (\ref{Psi}). Their another expression on $t$-axis, {i.e.,} $ z_i,~ {w}_i,~ \tau_{\theta_i},~ \tau_{\delta_{i}}$ and $ \Psi$  can be obtained immediately  according to Section \ref{System Reparameterization}. $\beta(\tau)=e^{\tau}/T$ and $\mu(t)=1/(T-t)$.   
	}\label{table 1}
\end{table*}
\begin{theorem}[\textbf{{\small Prescribed-time Control for Strict-Feedback Systems}}]\label{theorem1}
	Consider the closed-loop system consisting of the plant (\ref{system}), the prescribed-time adaptive controller and the  estimators as shown in TABLE \ref{table 1}. Under Assumptions \ref{assumption1} and \ref{assumption2}, the following results hold:
	\begin{itemize}
		\item[$1)$] All the closed-loop signals are globally bounded;
		\item[$2)$] The prescribed-time convergences of all system states and control input  are achieved, {i.e.,}  
		$
		\lim_{t\rightarrow T}x(t) =\lim_{t\rightarrow T}u(t)=0;
		$
		\item[$3)$]  $ \|\dot{\hat{\theta}}\|,~ \dot{\hat{\delta}}_\theta  $ and $\dot{\hat{\rho}}$ are uniformly bounded over $[0,T)$. Furthermore, $\lim_{t\rightarrow T}\hat{\theta}$, $\lim_{t\rightarrow T}\hat{\delta}_\theta$ and $\lim_{t\rightarrow T}\hat{\rho}$   exist  but they are not necessarily equal to $\ell_\theta$, $\delta_{\theta}$ and $1/\ell_b$.
	\end{itemize}  
\end{theorem}

\begin{remark}
	It is easy to see that Theorem \ref{theorem1} extends the results in \cite{2021-TAC-huachanghun-prescribed-adaptive} for time-invariant systems to time-varying systems  through a spatiotemporal transformation (see Fig. \ref{fig1}) based method.   In particular, the system considered in this paper has an unknown time-varying control coefficient, which poses additional difficulties for the design of the corresponding control algorithm. Contribution with respect to state-of-the-art\cite{2018-yucelen-finite,2020-NYU-dynamic-high-gain,2020-yucelen-time-scaling}, the temporal-scale transformation is used for the first time in this paper to construct a smooth control scheme over $[0,T)$ to achieve prescribed-time stabilization of nonlinear systems with unknown time-varying control coefficients.
\end{remark}

\begin{theorem}[\textbf{Prescribed-time Control for Normal-Form Systems}]\label{theorem2} 
	Consider system (\ref{system}) with $\phi_{i}(\underline{x}_i)\equiv 0$, $i=1,\cdots,n-1$ under Assumptions \ref{assumption1} and \ref{assumption2}. If the control law and the  estimators are designed as $u=\hat{\rho}\bar{u}$, 
	\begin{equation}\label{u-normal}
	\bar{u}=-  \left(k\mu(t)+\frac{\hat{\delta}_\theta  \phi_n^{\top}\phi_n}{\sqrt{s^2\phi_n^{\top}\phi_n+\varepsilon^2}}+\frac{\psi^2}{\sqrt{s^2\psi^2+\varepsilon^2}}\right)s_n,
	\end{equation}
	and
	\begin{equation} \label{update-normal}
	\begin{aligned}
	\dot{\hat{\theta}}&=\Gamma \phi_n s_n,~~~~~~~~~~~~~~~\hat{\theta}(0)\geq 0,~\Gamma\succ 0\\ 
	\dot{\hat{\delta}}_\theta&=\frac{\gamma_\delta s_n^2\phi_n^\top\phi_n}{\sqrt{s_n^2\phi_n^{\top}\phi_n+\varepsilon^2}},~~\hat{\delta}_\theta(0)\geq 0,~\gamma_{\delta}\geq 0,~\varepsilon>0\\ 		  
	\dot{\hat{\rho}}&=-\gamma_\rho\operatorname{sgn}(\ell_{b}) s_n\bar{u},~~ ~\gamma_{\rho}>0\\
	\end{aligned}
	\end{equation}
	with $s$ being the filter variable as defined in Lemma \ref{lemma-s}, $k>1/(\underline{b}\hat{\rho}(0))$, $\psi=\sum_{i=1}^{n}l_i\mu^{n-i+1} x_i+\phi_n^{\top}\hat{\theta}$, $\mu(t)=1/(T-t)$, and $\hat{\rho}(0)$ being constant chosen such that $\hat{\rho}(0)\operatorname{sgn}(\ell_{b})>0$, then the control objectives $1)-3)$ as stated in Theorem \ref{theorem1} are achieved. 
\end{theorem}

\begin{remark}
	Thanks to the non-regressor based design approach,   the state variables $x_i$ and the filter variable $s$ do not need to satisfy the one-to-one mapping relationship, so for normal form nonlinear systems, we can employ the classical filter variable $s$ in control design to avoid the explosion of computational complexity in the backstepping design\cite{krstic,chenkaiwen,chenkaiwenIFAC}, thus simplifying the controller structure  while reducing the computational cost of the algorithm without losing control accuracy.
\end{remark} 

\begin{remark}[\textbf{Implementation}]\label{implementation}
	It can be seen that Theorems \ref{theorem1} and \ref{theorem2} are well established over $[0,T)$. This is important and good enough in certain applications (e.g., missile guidance) that only need to operate for a finite time  interval. However, in order to ensure that the system converges to the equilibrium point within the prescribed time and that the system maintains equilibria everywhere in the state space past that time, we need to update the original controller using a non-stop running implementation, which is in fact quite simple, as follows
	\begin{equation} \label{new controller}
	u=\left\{\begin{array}{ll}
	\hat\rho \bar u, ~~~~~~t\in[0,T)\\   
	0,~~~~~~~\operatorname{otherwise}.
	\end{array}\right.
	\end{equation}
	Without considering external non-vanishing disturbances, this new controller (\ref{new controller}) guarantees the stability of the closed-loop system on $[0, \infty)$, and its proof   is straightforward and therefore omitted.
\end{remark}

\subsection{Two Corollaries}
Although Theorems \ref{theorem1} and \ref{theorem2} hold for a  finite-time interval that can be pre-set by users freely irrespective of initial conditions and any other design parameter, one may naturally ask whether this result can be generalized to an infinite time interval to achieve exponential or super-exponential stabilization.  The answer is yes, and we will describe how this is done in Corollaries \ref{corollary3} and \ref{corollary4}. It is worth noting that works on exponential or super-exponential stabilization  have been solved by the second author and his coauthors using a time-varying feedback (state scaling) based  strategy (see \cite{Zhao} and \cite{2019-wangyujuan-auto-general}), so Corollaries \ref{corollary3} and \ref{corollary4} aim to give an alternative  scheme while extending the application of the algorithm to parameter-varying nonlinear systems.

\begin{corollary}[\textbf{Exponential Control for Normal-Form Systems}]\label{corollary3}
	Consider system (\ref{system}) with $\{\phi_{i}(\underline{x}_i)\}_{i=1}^{n-1}\equiv 0$ under Assumptions \ref{assumption1} and \ref{assumption2}. The controller and the parameter  estimators are designed to be the same as (\ref{u-normal}) and (\ref{update-normal}).  If $\varepsilon$ is chosen to  be a time-varying function that satisfies $\int_{0}^{\infty}\varepsilon(t) dt \in\mathcal{L}_\infty$ and the function  $\mu(t)$  is replaced by $a'(t)$, where $a'(t)$ is derived from Definition \ref{definition-2}, then the closed-loop system is globally exponentially stable, i.e., $i)$ All the closed-loop signals are globally bounded; $ii)$ All system states can converge at an exponential rate to zero.
\end{corollary}

\begin{corollary}[\textbf{{\small Super-Exponential Control for Normal-Form Systems}}]\label{corollary4}
	Consider system (\ref{system}) with $\{\phi_{i}(\underline{x}_i)\}_{i=1}^{n-1}\equiv 0$ under Assumptions \ref{assumption1} and \ref{assumption2}. The controller and the parameter estimators are designed to be the same as (\ref{u-normal}) and (\ref{update-normal}). If $\varepsilon$ is chosen to  be a time-varying function that satisfies $\int_{0}^{\infty}\varepsilon(t) dt \in\mathcal{L}_\infty$ and the function  $\mu(t)$  is replaced by  $a'(t)$, where  $a'(t)$ is derived from Definition \ref{definition-3}, then the closed-loop system is globally super-exponentially stable,  i.e., $i)$ All the closed-loop signals are globally bounded;  $ii)$ All system states can converge at a super-exponential\footnote{Refer \cite{2019-wangyujuan-auto-general} for the concept of
		super-exponential convergence.} rate to zero.   
\end{corollary}

\begin{definition}\label{definition-2}
	The temporal axis mapping ($\tau=a(t) \Leftrightarrow t=a^{-1}(\tau)$) with the following properties is called an exponential-type of temporal-scale transformation\footnote{An example of exponential-type of temporal-scale transformation is $\tau=e^t-1\Leftrightarrow t=\ln (\tau+1)$. In this case,  $a'(t) =e^t$.}:
	\begin{itemize}
		\item $a(0) = 0$ and $a(\infty)=\infty$;
		\item $a(t)$ is continuously differential on $t\in[0,\infty)$;
		\item  $e^{\lambda_1t}\leq a'(t)\leq e^{\lambda_2 t} $, where $\lambda_1$ and $\lambda_2$ are constants and satisfy $0<\lambda_1<\lambda_2$.
	\end{itemize}
\end{definition}

\begin{definition}\label{definition-3}
	The temporal axis mapping ($\tau=a(t) \Leftrightarrow t=a^{-1}(\tau)$) with the following properties is called a super-exponential-type of temporal-scale transformation\footnote{An example of super-exponential-type of temporal-scale transformation is $\tau=\int_{0}^{t}\exp(\exp(v))dv+C$ with $C$ being a proper constant. In this case, $a'(t)=\exp(\exp(t))$.}:
	\begin{itemize}
		\item $a(0) = 0$ and $a(\infty)=\infty$;
		\item $a(t)$ is continuously differential on $t\in[0,\infty)$;
		\item  $\exp(\lambda_2\exp(\lambda_1t)) \leq a'(t)\leq \exp(\lambda_j\cdots(\exp(\lambda_2\exp(\lambda_1t)))) $, where  $0<j<\infty$, and $\lambda_1,\lambda_2,\cdots,\lambda_j$ are  positive constants.
	\end{itemize}
\end{definition}

\begin{remark}\label{remark9}
	Note that the algorithm in Theorem \ref{theorem2} ensures that the closed-loop system is asymptotically stable when we choose $\mu(t)\equiv1$. The proof is omitted since it is straightforward with a slight modification of the proof of Theorem \ref{theorem1}.  Consequently, with the results as stated  in Theorem \ref{theorem2}, Corollaries \ref{corollary3} and \ref{corollary4}, we establish a uniform control framework allowing the closed-loop system (\ref{xn}) to be regulated to zero asymptotically, exponentially, super-exponentially or within prescribed time.
\end{remark}

\subsection{Stability Analysis}
\subsubsection*{\textbf{Proof of Theorem \ref{theorem1}}}   Note that $\beta(\tau)=\frac{1}{T}e^{\tau}$. Thus, it follows that
\begin{equation} \label{epsilon}
\int_{0}^{\infty} \frac{(n\delta_{{\theta}}+1) {\varepsilon} }{\beta} d\tau = {(n\delta_{{\theta}}+1) {\varepsilon}} T(1-e^{-\tau})<\infty.
\end{equation}
By integrating the left
and right sides of (\ref{DVn}) on $[0,\tau)$, we obtain 
\begin{equation}\label{Vn}
\begin{aligned}
V  (\tau) 
\leq  V  (0)-\int_{0}^{\tau} k V_z dv + \int_{0}^{\tau} \frac{(n\delta_{{\theta}}+1) {\varepsilon} }{\beta} dv <\infty ,
\end{aligned}
\end{equation} 
then,
\begin{equation} \label{int_Vz}
\begin{aligned}
\int_{0}^{\tau} k V_z dv
\leq  V  (0) + \int_{0}^{\tau} \frac{(n\delta_{{\theta}}+1) {\varepsilon} }{\beta} dv <\infty ,
\end{aligned}
\end{equation}
where $k=2\min\{k_1,k_2,\cdots,k_n\}$. It follows from (\ref{47}), (\ref{Vn}) and (\ref{int_Vz}) that $V,V_z,V_\theta,V_\rho,V_\delta$ and $\breve{z}_i$ are bounded for $\forall \tau\in[0,\infty)$. Furthermore,  by rewriting (\ref{DVn})  as $\frac{d V}{d\tau}\leq -kV +k(V_\theta+V_\rho+V_\delta)   +\frac{(n\delta_{{\theta}}+1) {\varepsilon}}{\beta}$, we can solve the differential inequality (\ref{DVn}) via the well-known method of  \textit{variation of constants}  [\citen{ODE} \textit{Chap. IV}], resulting in 
\begin{equation} \label{Vz}
\begin{aligned}
V (\tau)\leq &e^{-k\tau} V(0)+e^{-k\tau} \int_{0}^{\tau} e^{kv} \frac{(n\delta_{{\theta}}+1) {\varepsilon} }{\beta} dv +e^{-k\tau} \\
&\times \int_{0}^{\tau}ke^{kv}(V_\theta+V_\rho+V_\delta) dv.
\end{aligned}
\end{equation}
Since the second line of (\ref{Vz}) can be   rewritten using the   method of integration by parts as the following equation:
\begin{equation} \label{integration by parts}
\begin{aligned}
\int_{0}^{\tau}ke^{k\tau}&(V_\theta+V_\rho+V_\delta) ds= e^{k\tau}(V_{\theta}+V_\rho+V_\delta)\\
&-\int_{0}^{\tau}e^{kv}\left(\frac{dV_\theta}{d\tau}+\frac{dV_\rho}{d\tau}+\frac{dV_\delta}{d\tau}\right)dv,
\end{aligned}
\end{equation}
then, 
\begin{equation} \label{Vz-z}
\begin{aligned}
V_z (\tau)\leq &e^{-k\tau} V(0)+e^{-k\tau} \int_{0}^{\tau} e^{kv} \frac{(n\delta_{{\theta}}+1) {\varepsilon} }{\beta} dv \\
&-e^{-k\tau}\int_{0}^{\tau}e^{kv}\left(\frac{dV_\theta}{d\tau}+\frac{dV_\rho}{d\tau}+\frac{dV_\delta}{d\tau}\right)dv.
\end{aligned}
\end{equation}
Synthesize the above analysis, we apply L'Hospital's rule to $V_z$ to get
\begin{equation}  \label{bounded-DVi}
\begin{aligned}
\lim_{\tau\rightarrow\infty}V_z (\tau)&=  -\lim_{\tau\rightarrow\infty}\frac{\int_{0}^{\tau}e^{kv}\left(\frac{dV_\theta}{d\tau}+\frac{dV_\rho}{d\tau}+\frac{dV_\delta}{d\tau}\right)dv}{e^{ k\tau}}\\
&=-\frac{1}{k}\lim_{\tau\rightarrow\infty}\left(\frac{dV_\theta}{d\tau}+\frac{dV_\rho}{d\tau}+\frac{dV_\delta}{d\tau}\right) .
\end{aligned}
\end{equation}
\begin{remark}
	Our analysis is partly motivated by \cite{2021-TAC-huachanghun-prescribed-adaptive}, but here the analysis is performed on an infinite time domain and therefore does not rely on a specific lemma on the improper integral as involved in finite time domain. 
\end{remark}

\begin{table*}[!]
	\rule[1pt]{18cm}{0.05em} 
	\begin{equation}\label{compact-set}        
  \begin{aligned}
	\frac{d \breve z_1}{d\tau}=&-\left(k_1+\frac{{\hat{\delta}_\theta}   \breve{w}_1^{\top}\breve{w}_1}{\beta\sqrt{\breve{z}_1^2\breve{w}_1^{\top}\breve{w}_1+ \varepsilon^2 }}\right)\breve{z}_1+\frac{1}{\beta }\left(\breve{z}_2 +\breve w_1^{\top}(\theta(\tau)-\hat{\theta})\right) \triangleq -\mathsf{K}_1\breve{z}_1+\frac{1}{\beta}\mathcal{Y}_1
	\\ 
	\frac{d \breve z_i}{d\tau}=&-\left(k_i+\frac{{\hat{\delta}_\theta}   \breve{w}_i^{\top}\breve{w}_i}{\beta\sqrt{\breve{z}_i^2\breve{w}_i^{\top}\breve{w}_i+ \varepsilon^2 }}\right)\breve{z}_i+\frac{1}{\beta }\left[  \frac{\gamma_{\delta}   \breve{z}_i\breve{w}_i^{\top}\breve{w}_i}{\sqrt{\breve{z}_i^2\breve{w}_i^{\top}\breve{w}_i+ \varepsilon^2}}  \sum_{j=2}^{i-1}\frac{\partial \breve\alpha_{j-1}}{\partial \hat\delta_\theta}  \breve{z}_j  +\sum_{j=2}^{i-1}\frac{\partial \breve\alpha_{j-1}}{\partial \hat\theta}\Gamma_1\breve{w}_i\breve{z}_j -\breve{z}_{i-1}  +\breve z_{i+1} \right.\\ 
	& \left. +\breve w_i^{\top}(\theta(\tau)-\hat{\theta})+\frac{\partial \breve\alpha_{i-1}}{\partial \hat{\theta}}\left( {\Gamma_1} \breve\tau_{\theta_i}  -\beta\frac{\partial \hat{\theta}}{\partial \tau}\right)+\frac{\partial \breve\alpha_{i-1}}{\partial \hat{\delta}_\theta}\left( {\gamma_{\delta}} \breve{\tau}_{\delta_{i}}-\beta\frac{\partial \hat{\delta}_\theta}{\partial \tau}\right)\right]\triangleq  -\mathsf{K}_2\breve{z}_2+\frac{1}{\beta}\mathcal{Y}_2, ~~i=2,\cdots,n-1\\
	\frac{d \breve z_n}{d\tau}=&-\left(k_n  +  \frac{{\hat{\delta}_\theta}   \breve{w}_n^{\top}\breve{w}_n}{\beta\sqrt{\breve{z}_n^2\breve{w}_n^{\top}\breve{w}_n+ \varepsilon^2 }} \right)b(\tau)\hat{\rho}\breve{z}_n+\frac{1}{\beta }\left[b(\tau)\hat{\rho}\left(   -\frac{\breve\Psi^2\breve z_n}{\sqrt{\breve{z}_n^2\breve\Psi^2+ {\varepsilon}^2}}\right) -\sum_{j=1}^{n-1}\frac{\partial\breve{\alpha}_{n-1}}{\partial\breve{x}_j}\breve{x}_{j+1} \right. \\
	&  \left. -\beta \frac{\partial \breve\alpha_{n-1}}{\partial \beta }\frac{\partial \beta }{\partial \tau}+\breve w_n^{\top}  \theta(\tau)  -\beta\frac{\partial \breve\alpha_{n-1}}{\partial \hat{\theta}}\frac{\partial \hat{\theta}}{\partial \tau}-\beta\frac{\partial \breve\alpha_{n-1}}{\partial \hat{\delta}_\theta}\frac{\partial \hat{\delta}_\theta}{\partial \tau}   \right] \triangleq -\mathsf{K}_n\breve{z}_n+\frac{1}{\beta}\mathcal{Y}_n\\
	\end{aligned}         
	\end{equation} \\\rule[1pt]{18cm}{0.05em} 
\end{table*}
Now, to show the asymptotic convergence of $\breve{z}_i(\tau)$, we substitute virtual and actual control inputs into (\ref{open-loop}), one can express the dynamics of the closed-loop system as follows: 
\begin{equation}\label{closed-loop}
\begin{aligned}
\frac{d \breve z_i}{d\tau}=-\mathsf{K}_i(\tau) \breve z_i+\frac{1}{\beta} \mathcal{Y}_{i}(\underline{\breve z}_{i+1},\hat{\theta},\hat{\delta}_\theta,\tau),
\end{aligned}
\end{equation} 
where $\mathsf{K}_i(\tau)$ is a bounded function and satisfies $k_i=\inf\{\mathsf{K}_i(\tau)\}$,   and $\mathcal{Y}_i$ is a computable function. The  complete expression of (\ref{closed-loop}) is shown in (\ref{compact-set}), which, in combination with  (\ref{V_theta}), (\ref{V_rho}) (\ref{d_theta}), (\ref{d_rho})   (\ref{V_delta})  (\ref{update law of delta}) and (\ref{bounded-DVi}), implies that $\frac{1}{\beta}\mathcal{Y}_i\in \mathcal{L}_\infty$. Therefore, it can be  concluded that ${d\breve{z}_i}/{d\tau}\in\mathcal{L}_{\infty}$. Since (\ref{Vn}) and (\ref{int_Vz}) show that $\breve{z}_i\in\mathcal{L}_2\cap\mathcal{L}_\infty$, then using Barbalat's Lemma yields $\lim_{\tau\rightarrow\infty}\breve{z}_i=0$, which further indicates that $\lim_{\tau\rightarrow\infty}V_z=0$ and $\lim_{\tau\rightarrow\infty}({dV_\theta}/{d\tau}+{dV_\rho}/{d\tau}+{dV_\delta}/{d\tau})=0$.

To go on, we show the boundedness of $\breve\alpha_i$. By carefully examining   (\ref{compact-set}) and (\ref{closed-loop}), we see that  $\mathcal{Y}_1= \breve{z}_2 +\breve w_1^{\top}(\theta(\tau)-\hat{\theta})  $, $\mathcal{Y}_1\in\mathcal{C}^{\infty}$ and $ \mathcal{Y}_1(0)=0$. Also note that $\beta=e^{\tau}/T$ and $d\beta/d\tau=\beta$, so by Lemma \ref{lemma1}  we can select $k_1>\lim_{\tau\rightarrow\infty}\frac{1}{\beta}\frac{d\beta}{d\tau}\equiv 1$ such that $\lim_{\tau\rightarrow\infty}\breve{z}_1\beta =0$.  Therefore, the boundedness of $\breve{\alpha}_1$ can be guaranteed and it can be seen that $\lim_{\tau\rightarrow\infty}\breve{\alpha}_1=0$. 

Recalling $\lim_{\tau\rightarrow\infty}\breve{z}_1\beta=0$, $\lim_{\tau\rightarrow\infty}\breve{z}_i=0$ and $\lim_{\tau\rightarrow\infty}({dV_\theta}/{d\tau}+{dV_\rho}/{d\tau}+{dV_\delta}/{d\tau})=0$, then it follows from (\ref{compact-set}) and (\ref{closed-loop}) that
\begin{equation} 
\begin{aligned}
\frac{d \breve z_2}{d\tau}=-\mathsf{K}_2(\tau) \breve z_2+\frac{1}{\beta} \mathcal{Y}_{2}(\underline{\breve z}_3,\hat{\theta},\hat{\delta}_\theta,\tau),
\end{aligned}
\end{equation} 	
where $\mathcal{Y}_2\in\mathcal{C}^{\infty}$ and $\mathcal{Y}_2(0,\hat{\delta}_\theta,\hat{\theta},\tau)=0$. According to Lemma \ref{lemma1}, the boundedness of $\breve{z}_2\beta$ and the fact $\lim_{\tau\rightarrow\infty}\breve{z}_2\beta=0$ can be guaranteed by choosing $k_2>1$, thereby the boundedness of $\breve\alpha_2$ as well as the fact $\lim_{\tau\rightarrow\infty}\breve\alpha_2=0$ can be guaranteed. 

Furthermore, the fact of $\lim_{\tau\rightarrow\infty}\breve{z}_1\beta = \lim_{\tau\rightarrow\infty}\breve{z}_2\beta=0$  means   that we can rewrite (\ref{closed-loop}) as 
\begin{equation}\label{closed-loop-1}
\begin{aligned}
\frac{d \breve z_1}{d\tau}=-\mathsf{K}_1(\tau) \breve z_1+\frac{1}{\beta^2} \mathcal{Y}_{1,1}(\underline{\breve z}_1,\hat{\theta},\hat{\delta}_\theta,\tau),
\end{aligned}
\end{equation}
where $\mathcal{Y}_{1,1}=\beta\mathcal{Y}_1= \beta\breve{z}_2 +\beta\breve w_1^{\top}(\theta(\tau)-\hat{\theta})$ and satisfies that $\mathcal{Y}_{1,1}\in\mathcal{C}^{\infty}$ and $ \mathcal{Y}_{1,1}(0)=0$. Therefore, it follows from Lemma \ref{lemma1} that $\lim_{\tau\rightarrow\infty}\breve{z}_1\beta^2=0$ holds for $k_1>2$. 

Similarly, by utilizing  $\lim_{\tau\rightarrow\infty}\breve{z}_1\beta^2=0$, $\lim_{\tau\rightarrow\infty}\breve{z}_2\beta=0 $  $\lim_{\tau\rightarrow\infty}\breve{z}_i=0$ and $\lim_{\tau\rightarrow\infty}({dV_\theta}/{d\tau}+{dV_\rho}/{d\tau}+{dV_\delta}/{d\tau})=0$, we prove that $\mathcal{Y}_3$ is smooth and $\mathcal{Y}_3(0,\hat{\delta}_\theta,\hat{\theta},\tau)=0$. Thus, we can choose $k_3 > 1$ to prove that $\lim_{\tau\rightarrow\infty}\breve{z}_3\beta=0$, $\breve\alpha_3\in\mathcal{L}_\infty$ and $\lim_{\tau\rightarrow\infty}\breve{\alpha}_3=0$ with the support of Lemma \ref{lemma1}. 
By analogy, by selecting $k_1>3$ and $k_2>2$, we can prove that  $\lim_{\tau\rightarrow\infty}\breve{z}_1\beta^3=0$ and $\lim_{\tau\rightarrow\infty}\breve{z}_2\beta^2=0$, which further indicates $\lim_{\tau\rightarrow\infty}\breve{\alpha}_4=0$. Finally,  following the argument similar to the previous paragraph, we can conclude that    $\lim_{\tau\rightarrow\infty}\breve{z}_i\beta^{n-i+1}=0$ holds when we selecting   $k_i>n-i+1,~(i=1,\cdots,n-1)$. Therefore, the boundedness of   $\breve\alpha_i$ can be guaranteed and $\lim_{\tau\rightarrow\infty}\breve\alpha_i=0$ holds. 

Subsequently, it is straightforward to prove that $\lim_{\tau\rightarrow \infty}\breve{x}(\tau)=0$ by taking the transformation as defined in (\ref{error variables}).
Note that in the $n$-th step, the minimum value of $\mathsf{K}_n$ is $k_n\underline{b}\hat{\rho}(0)$ rather than $k_n$. Hence, we need to select $k_n\underline{b}\hat{\rho}(0)>1$, namely  $
k_n>\frac{1}{\underline{b}\hat{\rho}(0)},$
such that    $\breve{\bar u}\in\mathcal{L}_{\infty}$  and $ \lim_{\tau\rightarrow\infty}\breve{\bar u}=0$. These results  further indicate  that $\lim_{t\rightarrow T} \alpha_i(t)=\lim_{t\rightarrow T}\bar u(t)=0$. 
In addition, it follows from  $V_\rho\in\mathcal{L}_\infty$ and $u(t)=\hat{\rho}\bar{u}$ that $\hat{\rho}(t)\in \mathcal{L}_{\infty}$ and $\lim_{t\rightarrow T}  u(t)=0$.

The prescribed-time convergence of $\hat{\theta}(t)$, $\hat{\delta}_\theta(t)$ and $\hat{\rho}(t)$ can be obtained by proving the asymptotic convergence of $\hat{\theta}(\tau)$, $\hat{\delta}_\theta(\tau)$ and $\hat{\rho}(\tau)$ on $[0,\infty)$. Since $\breve{w}_i(0,\hat{\theta},\hat{\delta}_\theta,\tau)\equiv0$ and $\breve{w}_i$, $\hat{\delta}_\theta$ and $\hat{\theta}$  are bounded functions, it follows from (\ref{d_theta}), (\ref{d_rho}) and (\ref{update law of delta}) that there exists a number $L$ such that 
\begin{equation}\label{update-laws}
\begin{aligned}
&\| {d\hat{\theta}}/{d\tau}\|\leq L \|\breve{z}\|^2,~\\
&| {d\hat{\delta}_\theta}/{d\tau}|\leq   L \|\breve{z}\|^2,~\\
&|d\hat{\rho}/d\tau|\leq L \|\breve{z}\|^2.
\end{aligned}
\end{equation}
Therefore, $ \|\dot{\hat{\theta}}\|,~ \dot{\hat{\delta}}_\theta  $ and $\dot{\hat{\rho}}$ are uniformly bounded over $[0,T)$. Since $z\in\mathcal{L}_2$, then $\frac{d\hat{\theta}}{d\tau}\in\mathcal{L}_1$, $\frac{d\hat{\delta}_\theta}{d\tau}\in\mathcal{L}_1$, and $\frac{d\hat{\rho}}{d\tau}\in\mathcal{L}_1$. It follows from the argument similar to Theorem 3.1 in \cite{Krstic1996} 
that $\hat{\theta}(\tau)$, $\hat{\delta}_\theta(\tau)$ 
and $\hat{\rho}(\tau)$ have a limit as 
$\tau\rightarrow\infty$. Namely, $\hat{\theta}(t)$, $\hat{\delta}_\theta(t)$ and $\hat{\rho}(t)$ converge to a constant within a prescribed-time. This completes the proof. $\hfill\blacksquare$

\subsubsection*{\textbf{Proof of Theorem \ref{theorem2}}} To start with, we rewrite the considered systems as 
\begin{equation}\label{xn}
x_1^{(n)}=b(t)u+\phi_n^{\top}\theta(t).
\end{equation}
By recalling (\ref{s}), it could be easily checked that $\{s_i\}_{i=1}^n$ are linear combination of $\{x_i\}_{i=1}^n$, hence the dynamics of $s_n$ becomes
\begin{equation}
\begin{aligned}
\dot{s}_n
= b(t)u+\sum_{i=1}^{n}\frac{l_{ i}x_{i}}{(T-t)^{n-i+1}}+\phi_n^{\top} {\theta}(t),
\end{aligned}
\end{equation}
where $ l_{ i }$ is some known design parameter  related to   $k_i$, $i=1,\cdots,n$. Subsequently, one can obtain
\begin{equation}
\begin{aligned}
\frac{ds_n}{d\tau} 
=\frac{1}{\beta}&\left(b(\tau)\breve u+\sum_{i=1}^{n}l_i\beta^{n-i+1}\breve{x}_{i}\right.\\
&\left.~~+\breve\phi_n^{\top}\hat{\theta}+\breve\phi_n^{\top}(\ell_{\theta}-\hat{\theta})+\breve\phi_n^{\top}\Delta_{\theta}\right) ,
\end{aligned}
\end{equation}
where  $\Delta_{\theta}=\theta(\tau)-\ell_{\theta}$. Then, the Lyapunov function is chosen by
\begin{equation}
\begin{aligned}
V=&\frac{1}{2}s_n^2+\frac{1}{2}(\ell_{\theta}-\hat{\theta})^{\top}\Gamma^{-1}(\ell_{\theta}-\hat{\theta})+\frac{|\ell_{b}|}{2\gamma_{\rho}}\left(\frac{1}{\ell_{b}}-\hat{\rho}\right)^2\\
&+\frac{1}{2\gamma_{\delta}}\left(\delta_{{\theta}}-\hat{\delta}_\theta\right)^2.
\end{aligned}
\end{equation}
Recall that $\breve{u}=\hat{\rho}\breve{\bar u}$ and $\{\phi_{i}(\underline{x}_i)\}_{i=1}^{n-1}=0$,  the derivative of $V$ \textit{w.r.t.} $\tau$ along the trajectory of (\ref{system}) is shown as
\begin{equation}
\begin{aligned}
\frac{d V}{d\tau}=&\frac{s_n}{\beta} \left(\breve{\bar u}+\sum_{i=1}^{n}l_i\beta^{n-i+1}\breve{x}_{i}+\breve\phi_n^{\top}\hat{\theta}+\breve\phi_n^{\top}\Delta_{\theta}\right)\\
&+\frac{1}{\beta}\Delta_{b}\hat{\rho}s_n\breve{\bar{u}}+ (\ell_{\theta}-\hat{\theta})^{\top}\Gamma^{-1}\left(\frac{\Gamma}{\beta}\breve\phi_ns_n-\frac{d\hat{\theta}}{d\tau}\right)\\
&-\frac{|\ell_{b}|}{ \gamma_{\rho}}\left(\frac{1}{\ell_{b}}-\hat{\rho}\right)\left(\frac{\gamma_{\rho}}{\beta}\operatorname{sgn}(\ell_{b})s_n\breve{\bar{u}}+\frac{d\hat{\rho}}{d\tau}\right) \\
&-\frac{1}{ \gamma_{\delta}}\left(\delta_{{\theta}}-\hat{\delta}_\theta\right) \frac{d{\hat{\delta}}_\theta}{d\tau},
\end{aligned}
\end{equation}
where $\Delta_{b}=b(t)-\ell_{b}$. Since $s_n\breve\phi_n^{\top}\Delta_{\theta}\leq \delta_{{\theta}}\varepsilon +\frac{\delta_{{\theta}}s_n^2\breve\phi_n^{\top}\breve\phi_n}{\sqrt{s_n^2\breve\phi_n^{\top}\breve\phi_n+\varepsilon^2}}$ and $s_n\breve\psi\leq \varepsilon + \frac{s_n^2\breve\psi^2}{\sqrt{s_n^2\breve\psi^2+\varepsilon^2}}$, then it follows from (\ref{beta}), (\ref{u-normal}), and (\ref{update-normal})  that
\begin{equation}\label{dVtau}
\small{ \begin{aligned}
	\frac{d V}{d\tau}\leq&\frac{s_n}{\beta} \left(\bar u +\sum_{i=1}^{n}l_i\beta^{n-i+1}\breve{x}_{i}+\breve\phi_n^{\top}\hat{\theta}+ \frac{\hat\delta_{{\theta}}s_n \breve\phi_n^{\top}\breve\phi_n}{\sqrt{s_n^2 \breve\phi_n^{\top}\breve\phi_n+\varepsilon^2}}\right) \\
	&+\frac{1}{\beta}\delta_{{\theta}}\varepsilon+\frac{1}{ \gamma_{\delta}}\left(\delta_{{\theta}}-\hat{\delta}_\theta\right) \left(\frac{\gamma_{\delta} s_n^2\breve\phi_n^{\top}\breve\phi_n}{\sqrt{s_n^2\breve\phi_n^{\top}\breve\phi_n+\varepsilon^2}}-\frac{d{\hat{\delta}}_\theta}{d\tau}\right)\\ 
	&+\frac{1}{\beta}\Delta_{b}\hat{\rho}s_n\breve{\bar{u}}\\
	\leq & -ks_n^2+\frac{1}{\beta}(\delta_{{\theta}}+1)\varepsilon+\frac{1}{\beta}\Delta_{b}\hat{\rho}s_n\breve{\bar{u}}.
	\end{aligned}}
\end{equation}
By selecting $\hat{\rho}(0)\operatorname{sgn}(\ell_{b})>0$, similar to (\ref{shedong}), one can prove that $\frac{1}{\beta}\Delta_{b}\hat{\rho}s_n\breve{\bar{u}}\leq 0$. Therefore, (\ref{dVtau}) becomes 
\begin{equation}\label{dVtau-n}
\begin{aligned}
\frac{d V}{d\tau}\leq  -ks_n^2+\frac{1}{\beta}(\delta_{{\theta}}+1)\varepsilon .
\end{aligned}
\end{equation}
Now, following the same argument used in the proof of Theorem \ref{theorem1}, we know that $\lim_{\tau\rightarrow\infty}s_n(\tau)=0$. In addition, one can immediately prove that $\lim_{\tau\rightarrow\infty}\{s_i(\tau)\}_{i=1}^{n-1}=0$ according to Lemma \ref{lemma-s}. Then, from $s_1=x_1$, $x_2=\dot{x}_1$, $\beta=d\tau/dt$ and $s_2= {k_1}\beta s_1+\frac{ds_1}{d\tau}\frac{d\tau}{dt}$, we know that $d{s}_1/d\tau=- {k_1}  s_1,~k_1>n$ as $s_2=0$. Hence, it can be deduced with the help of Corollary \ref{corollary} that $\lim_{\tau\rightarrow\infty} {\beta^n}s_1(\tau)=\lim_{\tau\rightarrow\infty} {\beta^n}\breve x_1(\tau)=0$. Repeating the above steps, one can continue to get, for $i=1,\cdots,n-1$, $\lim_{\tau\rightarrow\infty} {\beta^{n-i+1}}s_i(\tau)=0$. Based upon these results, we can proceed to prove the asymptotic convergence of $\breve{x}(\tau)$ to zero as $\tau\rightarrow \infty$ by exploiting the converging-input converging-output property of the filter variable $s_n$. In addition, it is easy to prove that $\lim_{t\rightarrow T}x(t)=0$ by recalling the principle of temporal-scale transformation. 

In view of Lemma \ref{lemma1}, the boundedness of $\beta(\tau) s_n$ can be guaranteed by selecting $k>1/(\underline{b}\hat{\rho}(0))$ since the closed-loop dynamics of $s_n$ can be written as
\begin{equation}
\frac{ds}{d\tau}=-\mathsf{K}(\tau) s+\frac{1}{\beta}\mathcal{Y}(x,\hat{\theta},\hat{\delta},\tau),~~k\hat{\rho}(0)\underline{b}=\inf\{\mathsf{K}(\tau)\},
\end{equation}
where $\mathcal{Y}(x,\hat{\theta},\hat{\delta},\tau)=0$ as $x=0$. Therefore, it follows that $\mu(t)s_n\in\mathcal{L}_\infty[0,T)$ and $\lim_{t\rightarrow T}\mu s_n=0$, which also indicate  that, for $i=1,\cdots,n$, $\mu^{n-i+1} x_i\in\mathcal{L}_\infty[0,T)$ and  $\lim_{t\rightarrow T}\mu^{n-i+1} x_i=0$,
establishing the same for $u(t)$.

Finally, the prescribed-time convergence of $\hat{\theta}(t)$, $\hat{\delta}_\theta(t)$ and $\hat{\rho}(t)$ can be guaranteed according to the same argument as used in the proof of Theorem \ref{theorem1}. This completes the proof. $\hfill\blacksquare$

\subsubsection*{\textbf{Proof of Corollary \ref{corollary3}}} 
Firstly, by modifying $\varepsilon$  to be a time-varying function that satisfies $\int_{0}^{\infty}\varepsilon(t) dt \in\mathcal{L}_\infty$, one can find that Eq. (\ref{epsilon}) still holds. Therefore, similar to the proof of Theorem \ref{theorem1}, it is not difficult to prove that the controller with the parameter estimators  given in (\ref{u-normal}) and (\ref{update-normal}) can stabilize system (\ref{xn}) asymptotically over $\tau\in[0,\infty)$.  Next, one can directly obtain that the controller with the parameter estimators given in (\ref{u-normal}) and (\ref{update-normal})  can stabilize system (\ref{system})  exponentially  over $t\in [0,\infty)$ with the help of  the temporal-scale transformation as defined in Definition \ref{definition-2}. In addition, the boundedness of all closed-loop signals  can be proved rigorously, as we did in the proof of Theorem \ref{theorem1}, where the detail process is omitted here due to space limit.   $\hfill\blacksquare$

\subsubsection*{\textbf{Proof of Corollary \ref{corollary4}}}  
The proof is omitted since it is similar to the proof 
of Corollary \ref{corollary3} and,   in fact,  is straightforward after completing the proof of Theorem \ref{theorem1}. $\hfill\blacksquare$


\section{SIMULATIONS}\label{Simulation}

In this section, two illustrative numerical examples are provided to verify the effectiveness of the main results. The first example is a benchmark example in the presence of time-varying parameters in the feedback and the input paths. The second example is a practical model obtained by the ``wing-rock" unstable motion.
\subsection*{Example 1: Benchmark} 
Consider the benchmark example adapted from \cite{krstic} as follows:
\begin{equation}
\begin{aligned}
&\dot x_1=x_2+\theta(t)x_1,~\\
&\dot x_2=x_3,~\\
&\dot x_3=b(t)u,
\end{aligned}
\end{equation}
with
\begin{equation*}
\begin{aligned}
b(t)&=1.4+0.2\cos(10t),\\
\theta(t)&=1+  0.6\cos(40x_1 t) +0.2\sin(x_3^2t) +0.2\operatorname{sgn}(\sin(20t)).
\end{aligned}
\end{equation*} 

Each of these parameters comprise of a constant nominal part and a time-varying part designed to destabilize the system. The lower bound of $b(t)$ is assumed to be known as $\underline{b}=1.2$, and the ``radius" of change of $\theta(t)$ is $\delta_\theta=1$, which is assumed to be unknown in our prescribed-time controller design. It can be verified that Assumptions \ref{assumption1} and \ref{assumption2} are satisfied. Consider now two controllers: Controller 1 is the prescribed-time controller proposed in Theorem \ref{theorem1}, and Controller 2 is the asymptotic controller proposed in [\citen{chenkaiwen}, Proposition 1]. For comparison, set the common design parameters as $k_1=k_2=k_3=6$, $\Gamma=\gamma_{\rho}=0.01$,   $\hat{\theta}(0)=0$, and $\hat{\rho}(0)=1$. For the parameters solely used in Controller 1, set $\gamma_{\delta}= 0.01$, $\varepsilon=0.1$, $\hat{\delta}_\theta(0)=0$, and $T=2s$. For the parameters solely used in Controller 2, set $\delta_{{\theta}}=1$. The initial condition is set to $[x_1(0);x_2(0);x_3(0)]=[0.2;0;-0.2]$.

\begin{figure} [!]
	\centering
	\includegraphics[height=4.2 cm]{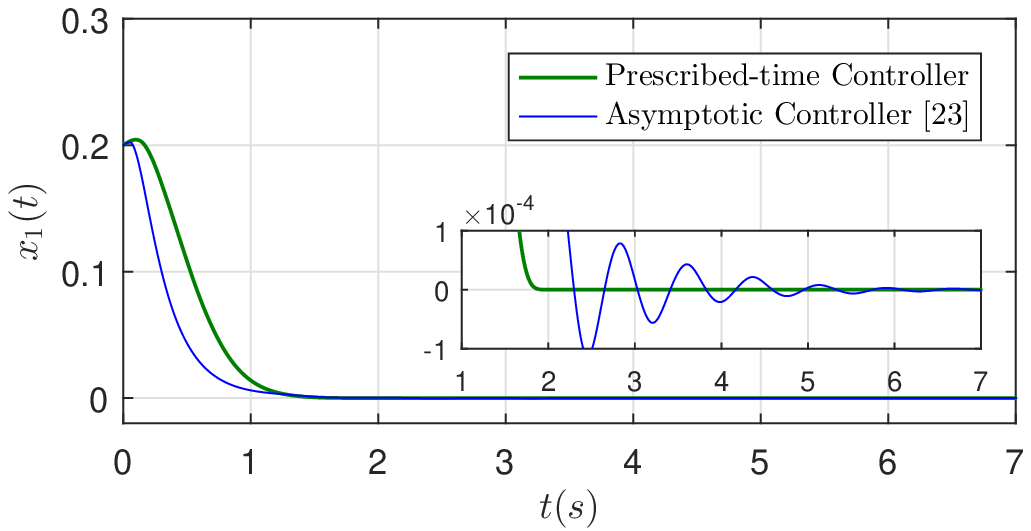}
	\includegraphics[height=4.2 cm]{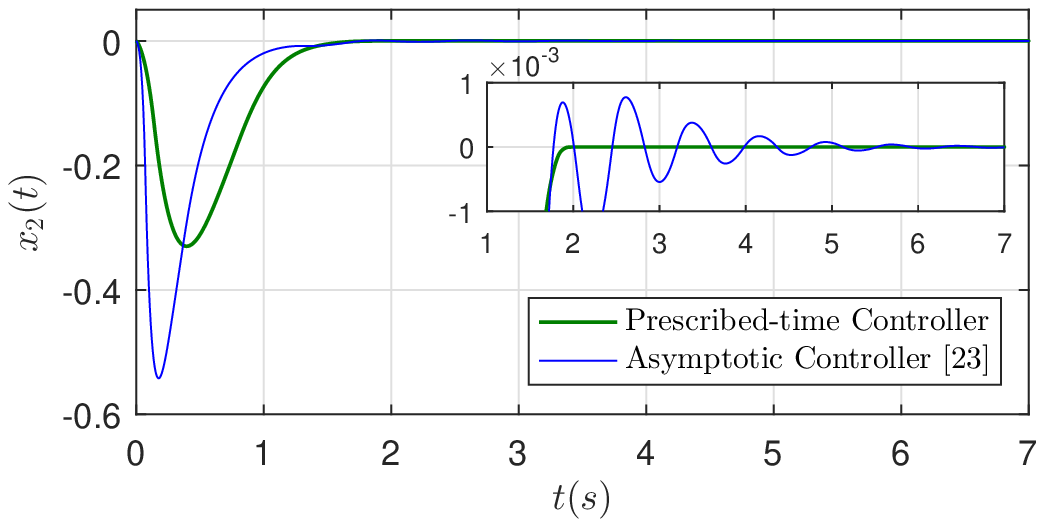}
	\includegraphics[height=4.2 cm]{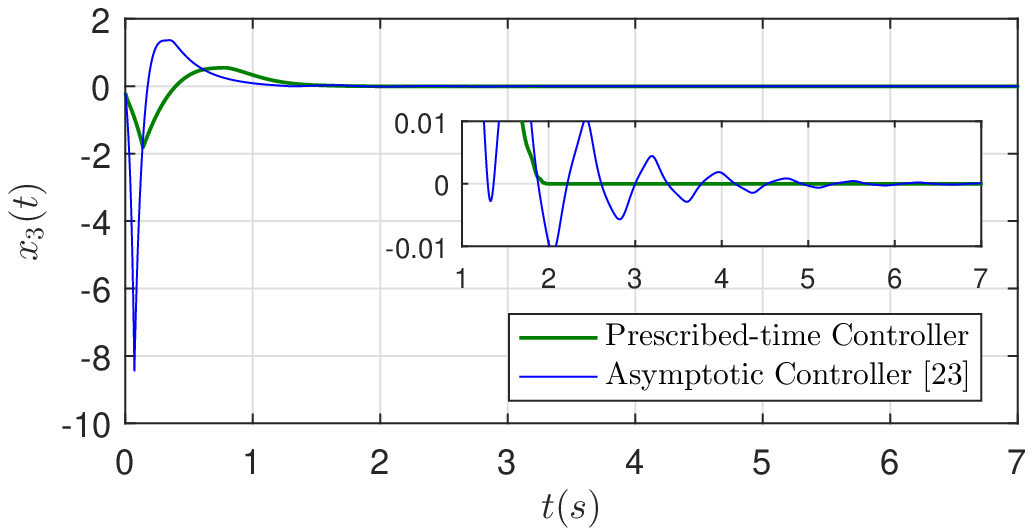}
	\caption{Trajectories of $x_i~(i=1,2,3)$  in $(t,x)$-plane for the initial condition $[x_1(0);x_2(0);x_3(0)]=[0.2;0;-0.2]$.}
	\label{fig1-benchmark} 
\end{figure}

\begin{figure} [!] 
	\centering
	\includegraphics[height=4.2 cm]{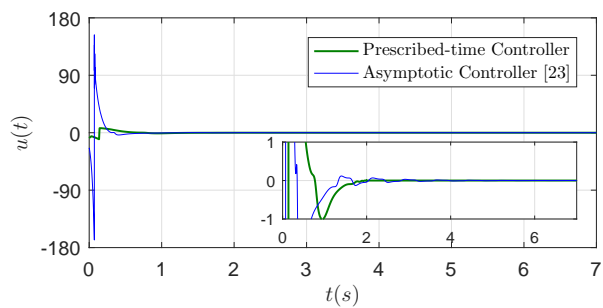}
	\caption{Trajectory of control input  $u(t)$.}
	\label{fig2-benchmark} 
\end{figure}

\begin{figure} [!]
	\centering
	\includegraphics[height=3.2cm]{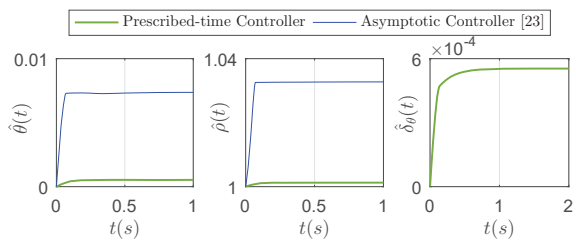}
	\caption{Trajectories of adaptive parameters $\hat\theta(t)$, $\hat\delta_{{\theta}}(t)$ and $\hat\rho(t)$.}
	\label{fig3-benchmark} 
\end{figure}

The simulation results are shown in Figs. \ref{fig1-benchmark}-\ref{fig3-benchmark}. From Figs. \ref{fig1-benchmark} and \ref{fig2-benchmark},  
we see that the system states, under Controller 1, are regulated to zero within the prescribed-time irrespective of initial condition and any other design parameter, and the control signals are continuous and steer to zero within the prescribed-time. It is also seen that the proposed control, as compared with that by \cite{chenkaiwen}, results in better transient and steady-state control performance with less control effort.  
This is partly due to the time-varying feedback introduced in the proposed algorithm, which gives the closed-loop system better transient performance, and partly due to the two-level adaptive estimation designed in Section \ref{Lyapunov Redesign with Tuning Functions}, which gives the algorithm a lower conservativeness. In addition, Fig. \ref{fig3-benchmark} show that  the corresponding adaptation parameters $\hat{\theta}(t)$, $\hat{\delta}_\theta(t)$ and $\hat{\rho}(t)$ converge ultimately to a non-zero constant. Furthermore, one can find that $\hat{\rho}(t)$ is a monotonically increasing function, which  confirms the theoretical analysis below (\ref{shedong}).



\subsection*{Example 2:  Model of ``Wing-rock" Unstable Motion}

Consider the scenario in which a high-performance airplane flying at high angle of attack aims at stabilizing its wing-rock unstable motion.  A single degree of freedom model is extracted from   \cite{Krstic1996model}, as follows
\begin{equation}\label{wink}
\begin{aligned}
&\dot \phi=p,~~\\
&\dot p= \frac{\bar q S b}{I_x}\left(0.5C_{l_1}\phi\sin(\alpha)+\frac{C_{l_2}pb}{2V}+C_{\delta_A}\right)
\end{aligned}
\end{equation}
where $\alpha$ is angle of attack in degrees, $\phi$ is the roll angle in radians, and $p$ is the roll rate in radians per second. The constants $\bar q, S, b, I_x$ and $V$ are the dynamic pressure, wing reference area, wing span, roll moment of inertia, and freestream air speed, respectively. The coefficients $C_{l_1}$ and $C_{l_2}$ are the rolling moment   derivatives, $C_{\delta_A}$ is the control surface.

The parametric strict-feedback form of the wing-rock model (\ref{wink}) by letting $x_1=\phi$, $x_2=p$ and $C_{\delta_A}=u$ is
\begin{equation}\label{simulation}
\begin{aligned}
&\dot x_1 =x_2+\phi_1^{\top}\theta(t),~\\
&\dot x_2 =b(t)u+\phi_2^{\top}\theta(t),
\end{aligned}
\end{equation}
where $\phi_1=0$, $\phi_2=[x_1,x_2]^{\top}$, $b(t)= {\bar q S b}/{I_x}$,  and 
$$\theta(t)= \left[\begin{array}{cccc}
\theta_1(t) \\  
\theta_2(t)  
\end{array}\right] = \left[\begin{array}{cccc}
0.5C_{l_1}\sin(\alpha)\bar{q}Sb/I_x \\  
C_{l_2}\bar q Sb^2/(2I_x V)
\end{array}\right] $$
Note that \cite{Krstic1996model} provides the following wind-tunnel data at angle of attack of $\alpha=30^{o}$: $\theta_1=-26.6667$ and $\theta_2=0.67485$. Taking into account that the change of the attack angle will cause $\theta$ to change, therefore we assume in the simulation that $\theta_1$ and $\theta_2$ will periodically change by $\pm 20\%$ on the basis of the experimental data, i.e., $\theta_i(t)=\theta_i+0.2\theta_i\operatorname{sgn}(\sin(3t))$ for $i=1,2.$  In addition, the high frequency gain is set as  $b(t)=2+0.2\operatorname{sgn}(\sin(3t))\cos(t)$. Note that except for its lower bound $\underline{b}=1.8$, 
the precise information on $b(t)$ is  unavailable (yet not needed)  for control design. For the system under consideration, it is readily verified that Assumptions \ref{assumption1}-\ref{assumption2}  are satisfied, thus the control schemes proposed in Remark \ref{remark9},  Corollaries \ref{corollary3}-\ref{corollary4}, as well as Theorem \ref{theorem2} can be directly applied to stabilize (\ref{simulation}) asymptotically, exponentially, super-exponentially and within prescribed time, respectively. All controllers and parameter estimators  share the same structure, as shown in (\ref{u-normal}) and (\ref{update-normal}), they differ only in the choice of some design parameters, as shown in Table \ref{table 2}. In addition, according to Lemma \ref{lemma-s}, we select the filter variable $s=k_1\mu(t) x_1+x_2$. For fair comparison, we set $[x_1(0);x_2(0)]=[0.2;0]$, $k_1=k=3$, $\hat{\theta}(0)=\hat{\delta}_\theta(0)=0$, $\hat{\rho}(0)=1$, $\gamma_\rho=\gamma_{\delta}=0.01$, and $\Gamma=I$ for all controllers. To ensure the prescribed-time controller share the property of non-stop running, an additional implementation scheme is used in the simulation as described in Remark \ref{implementation}.

\begin{table}[h]
	\caption{Parameter selection for different controllers}  
	\centering
	\begin{tabular}{c| c |c| c}
		\hline Controller & $\mu(t)$ & $\varepsilon$ & Convergence time\\
		\hline Asymptotic  & $1$ & $\exp(0.1t)$ &$\infty$ \\ 
		\hline Exponential & $\exp(t)$ & $\exp(0.1t)$ & $\infty$\\
		\hline Super-exponential  & $\exp(\exp(0.1t))$ & $\exp(0.1t)$ & $\infty$\\
		\hline  Prescribed-time & $1/(T-t)$ & $0.1$  & $T=0.5s$\\
		\hline
	\end{tabular}\label{table 2}
\end{table}

\begin{figure} [!]
	\begin{center}
		\includegraphics[height=4.2cm]{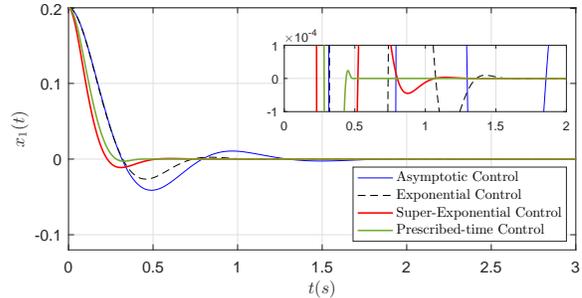}
		\caption{Trajectories of $x_1$ in $(t,x_1)$-plane under different controllers.}
		\label{fig2}
	\end{center}
\end{figure}

\begin{figure} [!]
	\begin{center}
		\includegraphics[height=4.2cm]{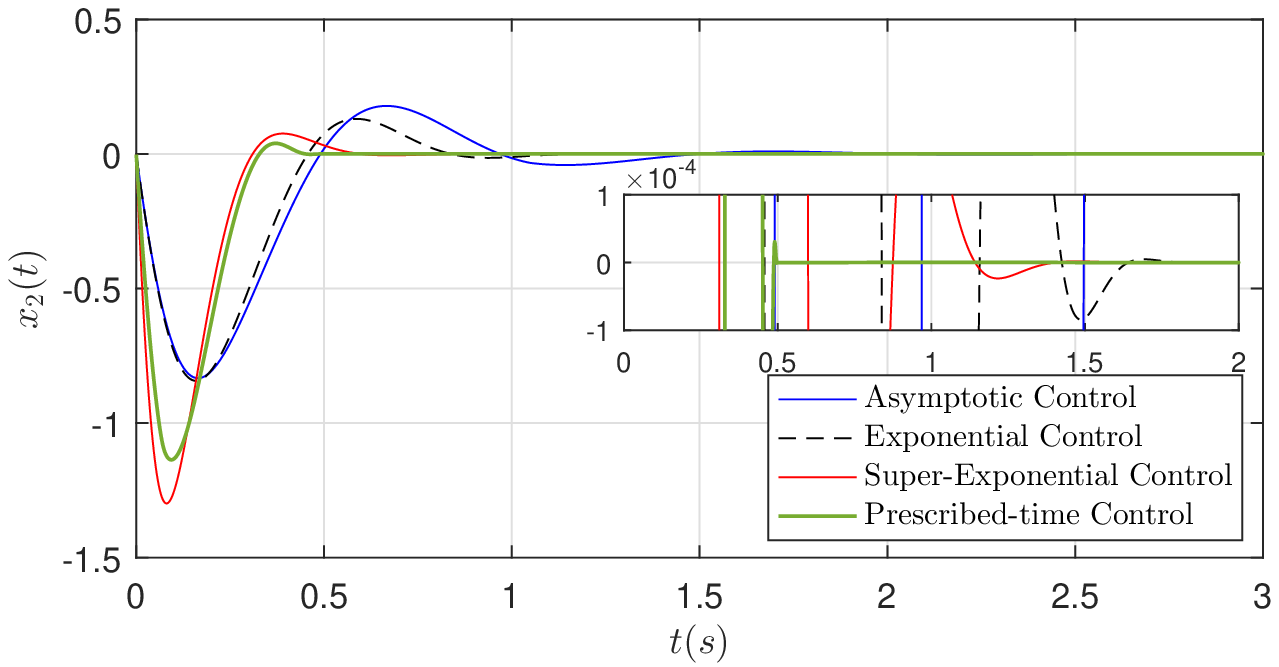}
		\caption{Trajectories of $x_2$ in $(t,x_2)$-plane under different controllers.}\label{fig3}
	\end{center}
\end{figure}

\begin{figure} [!]
	\begin{center}
		\includegraphics[height=4.2cm]{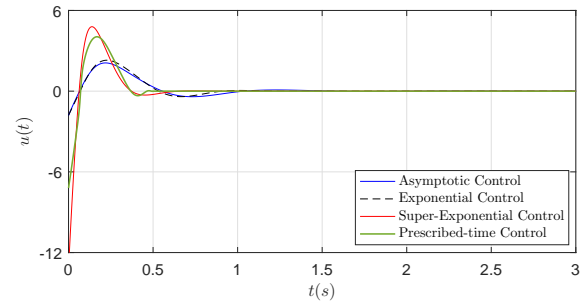}
		\caption{Trajectories of $u$ in $(t,u)$-plane under different controllers.}\label{fig4}
	\end{center}
\end{figure}

\begin{figure} [!]
	\begin{center}
		\includegraphics[height=6.2 cm]{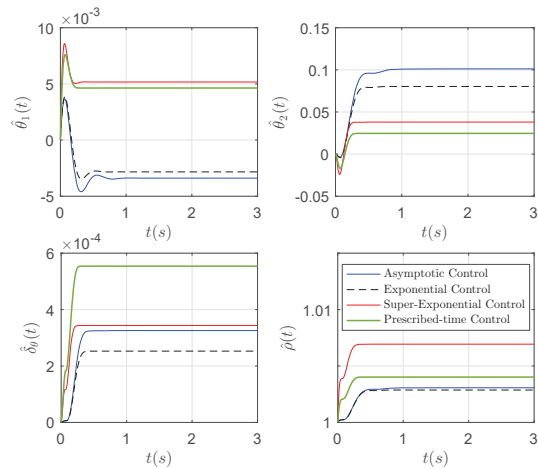}
		\caption{Trajectories of adaptive parameters in $(t,\hat{\theta}_1/\hat{\theta}_2/\hat{\delta}_\theta/\hat{\rho})$-plane under different controllers.}\label{fig5}
	\end{center}
\end{figure}

The responses of the state signals are shown in Figs. \ref{fig2}-\ref{fig3},   the responses of control input signals are shown in Fig. \ref{fig4}, and the evolutions of adaptive parameters are shown in Fig. \ref{fig5}.  From these simulation results, it is straightforward to see that prescribed-time convergence is faster than super-exponential convergence, super-exponential convergence is faster than exponential convergence, and exponential convergence is faster than asymptotic convergence. In addition,  it can be seen from Figs. \ref{fig2}-\ref{fig3} that the super-exponential  controller recovers the performance of the prescribed-time controller to some extent (i.e., it guarantees that all states converge to a small residual set within a short time). Furthermore, we see that the prescribed-time controller outperforms those infinite-time controllers since the settling time can be pre-set freely irrespective of the initial condition and design parameters. Finally, all results show that the proposed methods are powerful enough to stabilize the nonlinear system with fast time-varying parameters.


\section{CONCLUSIONS}\label{CONCLUSIONS}
This article presents a new adaptive prescribed-time stabilization method for parameter-varying nonlinear systems in strict-feedback form. Several new design techniques, e.g., spatiotemporal transformation, two-level estimation for fast time-varying parameters, and non-regressor based robust design, etc., are used in control design and stability analysis. By introducing a filtering variable based on the temporal-scale transformation, we develop a unified control framework for high-order nonlinear systems capable of achieving asymptotic, exponential, super-exponential, and prescribed-time convergence. It is interesting to note that with the proposed method  different convergence rates are realized with a unified control structure. 
Furthermore, unlike the related results about prescribed-time stabilization for systems with unknown control coefficients, where the selection of design parameters either relies on small-gain theorem\cite{2017-song-prescribed-time},  or on solving linear matrix inequalities\cite{2020-chitour-stabilization}, or on solving Lyapunov equation\cite{2021-shakouri-prescribed-linear-decay}, the selection of design parameters in this paper is guided by a novel Lemma, which is more concise and straightforward. In the simulation,  two illustrative numerical examples, a third-order benchmark example and a second-order practical model,  are provided to verify the benefits and effectiveness of the proposed schemes.  

One interesting topic for future study is to consider the output feedback adaptive prescribed-time control for high-order nonlinear systems by using state observers. Another research topic is to employ new tools (e.g., combining fractional power feedback and bounded time-varying gain\cite{2022-orlov-auto}) for  systems with unknown control coefficients and time-varying parameters to develop prescribed-time control schemes that can mitigate the effects of measurement noise.

\appendices
\section{Proof of Lemma \ref{lemma1}}
We begin with the solution to the homogeneous equation of (\ref{condition1}), and then perform the \textit{variation of constants} [\citen{ODE}, Chap. IV], as in 
\begin{equation}\label{x-solution}
\breve{z}_i(\tau)=C(\tau)e^{A(\tau)}
\end{equation}
where $A(\tau)=-\int_{0}^\tau \mathsf{K}_i(v)dv$. It follows that
\begin{equation}\label{dx=-ks+dc}
\begin{aligned}
\frac{d \breve{z}_i}{d\tau}&= -\mathsf{K}_i(\tau)C(\tau)e^{A(\tau)}+  \frac{dC(\tau)}{d\tau }e^{A(\tau )} .
\end{aligned}
\end{equation}
The solution that we are seeking should simultaneously satisfy the equation of motion—that follows from (\ref{condition1}) and (\ref{dx=-ks+dc}), namely,
\begin{equation}
\frac{d C(\tau)}{d\tau}=e^{ -A(\tau)}\frac{1}{\gamma^{\sigma}}\mathcal{Y}(\breve z_i,\tau).
\end{equation}
Inserting the initial condition $\breve{z}_i(0)$ into (\ref{x-solution}), we obtain a unique solution for (\ref{condition1}) as
\begin{equation}\label{xt}
\breve{z}_i(\tau)=e^{A(\tau)}\int_{0}^{\tau}e^{-A(v)}\frac{1}{\gamma^\sigma(v)}\mathcal{Y}(\breve{z}_i,v)dv+e^{A(\tau)}\breve{z}_i(0).
\end{equation}
To proceed, it follows from $0<k_i\leq  \mathsf{K}_i(\tau)$ that  
\begin{equation}
A(\tau) = -\int_{0}^\tau \mathsf{K}_i(v) dv\leq -\int_{0}^\tau k_i dv=-k_i\tau.
\end{equation} 
Recalling $\gamma(t)>0$ and applying condition (\ref{condition3}), we have
\begin{equation}
0\leq \lim_{\tau\rightarrow\infty} {e^{A(\tau)} }{\gamma^\sigma(\tau)}\leq \lim_{\tau\rightarrow\infty} {e^{-k_i\tau} }{\gamma^\sigma(\tau)}=0.
\end{equation}
Applying Squeeze Theorem, we obtain $\lim_{t\rightarrow T} {e^{A(\tau)}} {\gamma^\sigma(\tau)}=0$. Hence, 
\begin{equation}\label{x-eAt}
\lim_{\tau\rightarrow\infty} {e^{A(\tau)}{\gamma^\sigma(\tau)} \breve{z}_i(0)}=0.
\end{equation}
Using L'Hospital's rule on the basis of (\ref{xt}) and (\ref{x-eAt}), applying conditions (\ref{condition2}), (\ref{condition4}) and $\mathcal{Y}_i(0,\tau)=0$,  we have
\begin{equation}
\begin{aligned}
\lim_{\tau\rightarrow\infty} \breve{z}_i{\gamma^{\sigma}  }=&\lim_{\tau\rightarrow\infty}\frac{\int_{0}^{\tau}e^{-A(v)}\frac{1}{\gamma^\sigma }\mathcal{Y}_i(\breve{z}_i,v)dv}{e^{-A(\tau)} \gamma^{-\sigma} }\\
=&\lim_{\tau\rightarrow\infty}\frac{  \mathcal{Y}_i(\breve{z}_i,\tau) }{ \mathsf{K}_i(\tau) -    \frac{\sigma}{\gamma}   \frac{d\gamma }{d\tau}  } =0.
\end{aligned}
\end{equation}
This completes the proof. $\hfill\blacksquare$ 

\section{Proof of Lemma \ref{lemma-s}}
Solving the last differentiate equations in (\ref{s}) gives, 
\begin{equation}\label{96}
s_{n-1}(t)=e^{A(t)}s_{n-1}(0)+e^{A(t)}\int_{0}^te^{-A(v)}s_n(x,v)dv,
\end{equation}
where $A(t)=-\int_{0}^t\frac{k_{n-1}}{T-v}dv$. It is easy to check from (\ref{96}) that if $\int_{0}^{t}e^{-A(v)}s_n(x,v)dv$ is bounded, then $s_{n-1}(t)\rightarrow0$ as $t\rightarrow T$. If, however, $\int_{0}^{t}e^{-A(v)}s_n(x,v)dv$ is unbounded, we then  applying L'Hospital's rule to (\ref{96}) and obtain
\begin{equation} 
\lim_{t\rightarrow T}s_{n-1}(t)=0+\lim_{t\rightarrow T}  \frac{e^{-A(t)}s_n(x,t)(T-t)}{k_{n-1}e^{-A(t)}}=0
\end{equation}
which implies that $s_{n-1}$ converges to zero as $t\rightarrow T$. By carrying out the same procedure for the rest of the equations in (\ref{s}), one can conclude that   $\{s_i\}_{i=1}^{n-1}$ converges to zero within the prescribed-time $T$.  This completes the proof. $\hfill\blacksquare$

\begin{IEEEbiography}[{\includegraphics[width=1in,height=1.25in,clip,keepaspectratio]{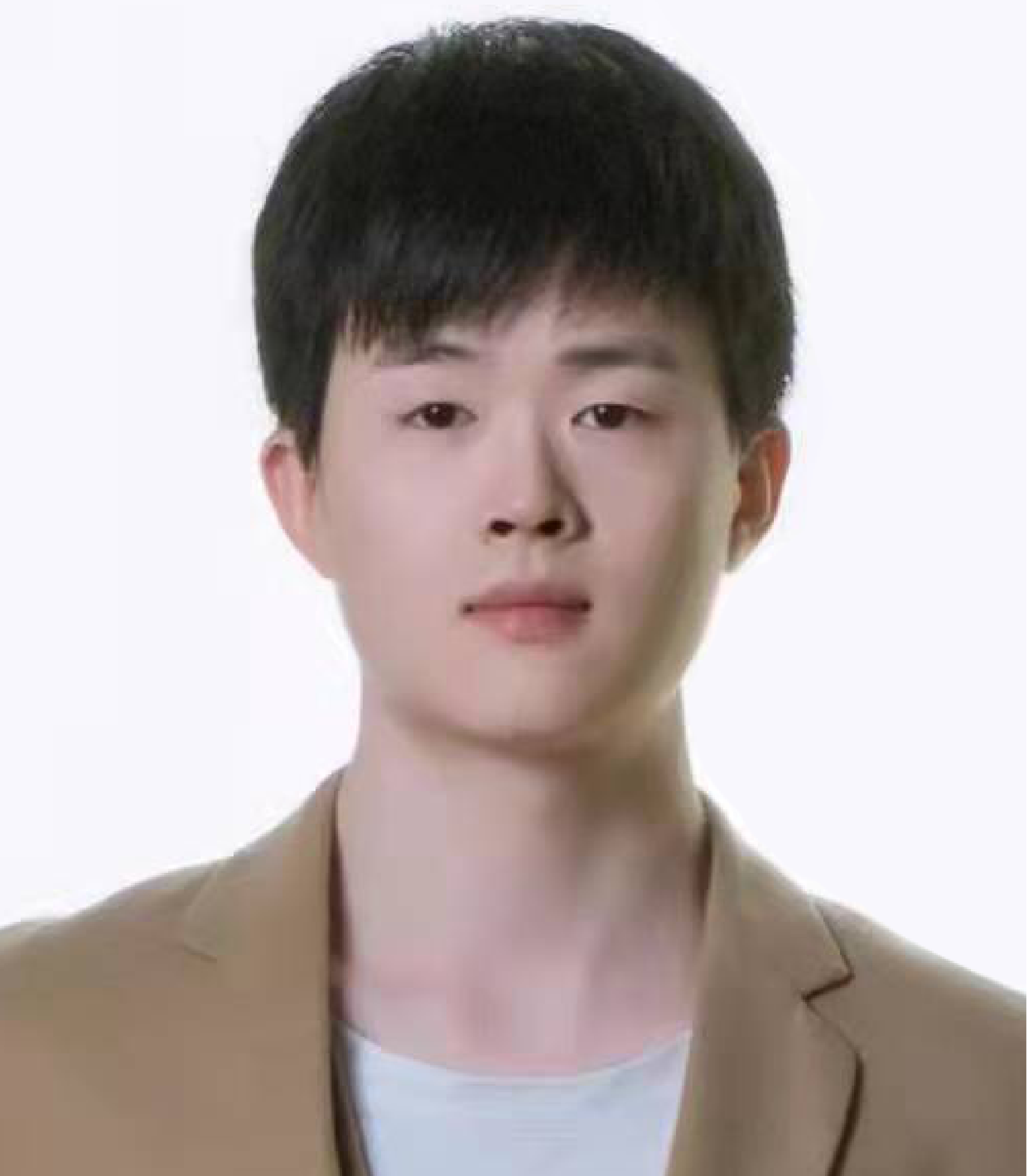}}]{Hefu Ye} received the B.Eng.  degree in the School of Information Science and Engineering in 2019 from Harbin Institute of Technology. He is currently pursuing the Ph.D. degree at the School of Automation, Chongqing University, 400044, China, and  he is now a Joint Ph.D. student at the School of Electrical and Electronic Engineering, Nanyang Technological University, 639798, Singapore. His research interests include multi-agent systems, robotic systems, robust adaptive control, prescribed performance control, and prescribed-time control. Dr. Ye is an active reviewer for many international journals, including the IEEE Transactions on Automatic Control, IEEE Transactions on Systems, Man, and Cybernetics: Systems, IEEE Transactions on Neural Networks and Learning Systems, etc.
\end{IEEEbiography}

 \begin{IEEEbiography}[{\includegraphics[width=1in,height=1.25in,clip,keepaspectratio]{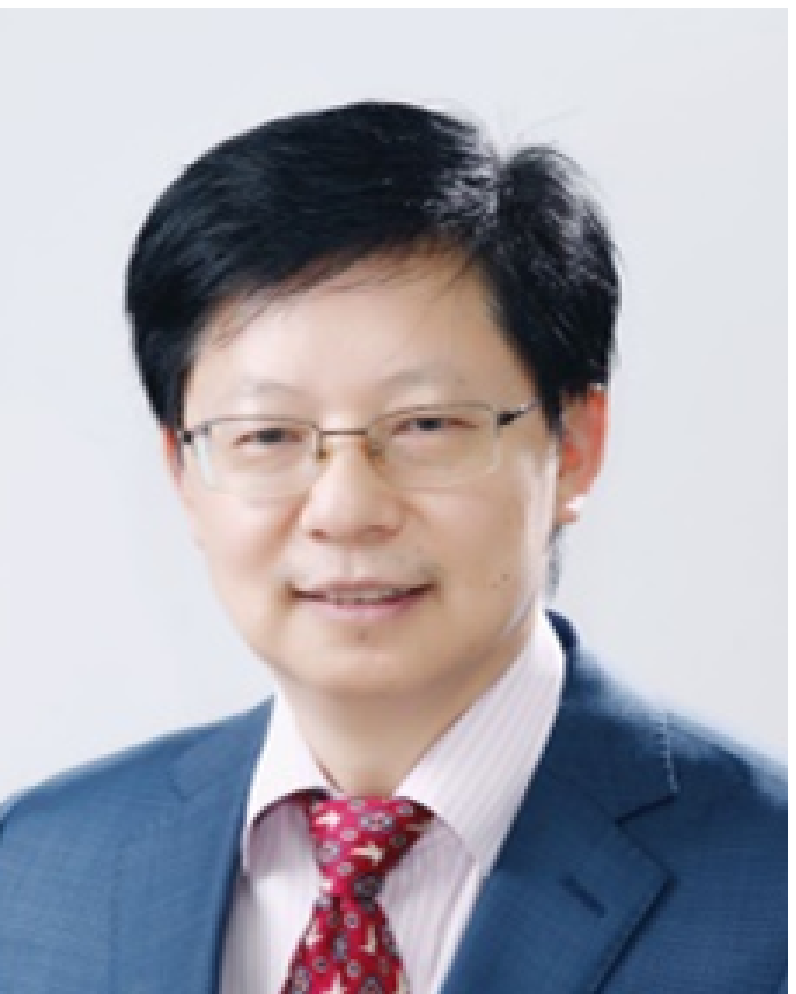}}]{Yongduan Song} (Fellow, IEEE) received the Ph.D. degree in electrical and computer engineering from Tennessee	Technological University, Cookeville, TN, USA, in 1992.	He held a tenured Full Professor with North Carolina A\&T State University, Greensboro, NC, USA, from 1993 to 2008 and a Langley Distinguished Professor	with the National Institute of Aerospace, Hampton, VA,	USA, from 2005 to 2008. He is currently the Dean of the School of Automation, Chongqing University, Chongqing, China. He was one of the six Langley 	Distinguished Professors with the National Institute of Aerospace (NIA), Hampton, VA, USA, and the Founding Director of Cooperative Systems with NIA. His current research interests include intelligent systems, guidance navigation and control, bio-inspired adaptive and cooperative systems. Prof. Song was a recipient of several competitive research awards from the National Science Foundation, the National Aeronautics and Space Administration, the U.S. Air Force Office, the U.S. Army Research Office, and the U.S. Naval Research Office. He is an IEEE Fellow and has served/been serving as an 	Associate Editor for several prestigious international journals, including the IEEE Transactions on Automatic Control, IEEE Transactions on Neural Networks and Learning Systems, IEEE Transactions on Intelligent Transportation Systems, IEEE Transactions on Systems, Man and Cybernetics, etc. 	He is currently Editor-in-Chief for the IEEE Transactions on Neural Networks and Learning Systems.  \end{IEEEbiography}

\end{document}